\documentclass[fleqn,usenatbib]{mnras}
\usepackage{newtxtext,newtxmath}
\usepackage[rgb]{xcolor}
\usepackage[T1]{fontenc}
\usepackage{ulem}
\usepackage{threeparttable}

\DeclareRobustCommand{\VAN}[3]{#2}
\let\VANthebibliography\thebibliography
\def\thebibliography{\DeclareRobustCommand{\VAN}[3]{##3}\VANthebibliography}

\def\gcc{\hbox{\rm\hskip.35em  g cm}$^{-3}$}

\def\radss{\hbox{\rm\hskip.35em  rad s}$^{-2}$}

\usepackage{graphicx}	
\usepackage{amsmath}	

\title[Multi-band study of pulsar glitches]{A multi-band study of pulsar glitches with \textit{Fermi}-LAT and Parkes}
\author[P. Liu et al.]{
P. Liu,$^{1,2}$
J.-P. Yuan,$^{2,3,4}$\thanks{E-mail: yuanjp@xao.ac.cn}
M.-Y. Ge,$^{5}$\thanks{E-mail: gemy@ihep.ac.cn}
W.-T. Ye,$^{3,5}$
S.-Q. Zhou,$^{6}$
S.-J. Dang,$^{7}$
Z.-R. Zhou,$^{8,9}$
 \newauthor
E. G{\"u}gercino{\u{g}}lu,$^{10}$ 
Z. H. Tu,$^{1}$
P. Wang,$^{10,11}$
A. Li,$^{1}$\thanks{E-mail: liang@xmu.edu.cn}
D. Li,$^{12,10,13}$
and N. Wang$^{2,4,14}$
\\
$^{1}$Department of Astronomy, Xiamen University, Xiamen, Fujian 361005, China\\
$^{2}$Xinjiang Astronomical Observatory, Chinese Academy of Sciences, 150 Science 1-Street, Urumqi, Xinjiang 830011, China\\
$^{3}$School of Astronomy, University of Chinese Academy of Sciences, Beijing 100049, China\\
$^{4}$Xinjiang Key Laboratory of Radio Astrophysics, 150 Science 1-Street, Urumqi, Xinjiang 830011, China\\
$^{5}$Key Laboratory of Particle Astrophysics, Institute of High Energy Physics, Chinese Academy of Sciences, Beijing 100049, China\\
$^{6}$School of Physics and Astronomy, Sun Yat-Sen University, Zhuhai 519082, China\\
$^{7}$School of Physics and Electronic Science, Guizhou Normal University, Guiyang 550001, China\\
$^{8}$National Time Service Center, Chinese Academy of Sciences, Xi'an 710600, China\\
$^{9}$Key Laboratory of Time Reference and Applications, Chinese Academy of Sciences, Xi'an 710600, China\\
$^{10}$National Astronomical Observatories, Chinese Academy of Sciences, 20A Datun Road, Chaoyang District, Beĳing 100101, China\\
$^{11}$Institute for Frontiers in Astronomy and Astrophysics, Beijing Normal University, Beijing 102206, China\\
$^{12}$Department of Astronomy, Tsinghua University, Beijing 100084, China\\
$^{13}$Computational Astronomy Group, Zhejiang Laboratory, Hangzhou 311100, China\\
$^{14}$Key Laboratory of Radio Astronomy, Chinese Academy of Sciences, Urumqi, Xinjiang 830011, China
}
\date{Accepted XXX. Received YYY; in original form ZZZ}
\pubyear{2024}

\begin{document}
\label{firstpage}
\pagerange{\pageref{firstpage}--\pageref{lastpage}}
\maketitle

\begin{abstract}
Pulsar glitch is a phenomenon characterized by abrupt changes in the spin period over less than a minute.
We present a comprehensive analysis of glitches in four gamma-ray pulsars by combining the timing observation data from \textit{Fermi} Large Area Telescope (\textit{Fermi}-LAT) and Parkes 64 m radio telescope. 
The timing data of five pulsars, namely PSRs J1028$-$5819, J1420$-$6048, J1509$-$5850, J1709$-$4429 (B1706$-$44) and J1718$-$3825, are examined over 14 yr of observations for each.
A total of 12 glitches are identified in four pulsars, including a previously unreported glitch.
That is, a new small glitch is identified for PSR J1718$-$3825 in MJD $\sim$ 59121(8), with a fractional glitch size of $\Delta \nu/\nu \sim 1.9(2) \times 10^{-9}$.
For PSR J1420$-$6048, our investigation confirms the presence of two linear recovery terms during the evolution of $\dot{\nu}$ following glitches 4, 6 and 8.
Moreover, an exponential recovery process was identified after glitch 8, with a recovery fraction ($Q$) of $Q = 0.0131(5)$ and a corresponding timescale of $\tau_{\rm d} = 100(6)$ d.
Regarding the fourth glitch of PSR J1709$-$4429, our analysis reveals the presence of two exponential recovery terms with degree of recovery and decay time-scales $Q$1 = 0.0104(5), $\tau_{\rm d1}=72(4)$ d and $Q$2 = 0.006(1), $\tau_{\rm d2}=4.2(6)$ d, respectively.
For the remaining previously reported glitches, we also refine the glitch epochs and recovery process through precise fitting of the timing data. 
We discuss how multi-band data of glitches can help better characterize the glitch recoveries and constrain the underlying physics of glitch events.
Our findings demonstrate that the accumulation of observational data reveals the rich complexity of the glitch phenomenon, aiding in the search for a well-established interpretation.
\end{abstract}


\begin{table*}  
\centering
\caption{Parameters and data spans for five gamma-ray pulsars studied in this paper. 
}    \label{Tab:publ-works}
\begin{tabular}{cccccccccc}
  \hline  \hline
PSR   & RA       & DEC & $P$  & $\dot{P}$  & $B_{\rm s}$  & $\tau_{\rm c}$ & DM &  \textit{Fermi}-LAT span  & Parkes span\\      &(hh:mm:ss)& ($\ast^{\degr} : \ast^{\arcmin} : \ast^{\arcsec}$) & (s) & (10$^{-15}$)   & ($10^{12}$ G) & (kyr) & (cm$^{-3}$pc) & (MJD)  & (MJD)\\  
\hline      
J1028$-$5819$^{a}$  &10:28:27.9(1)    &$-$58:19:06.21(6)  &0.091403  &16.1000  &1.23  &90.0   &96.525(2)     &54687--59809   &55233--59182 \\  
J1420$-$6048$^{b}$  &14:20:08.237(16) &$-$60:48:16.43(15) &0.068180  &83.1670  & 2.41  &13.0  &360.15(6)    &54722--59453   &54303--59182 \\ 
J1509$-$5850$^{c}$  &15:09:27.156(7)  &$-$58:50:56.01(8)  &0.088925  &9.1663   &0.91   &154.0   &142.1(1)      &54713--59510   &54303--59182 \\ 
J1709$-$4429$^{d}$  &17:09:42.728(2)  &$-$44:29:08.24(6)  &0.102459  &92.9845  &3.12   &17.5  &75.593(3)     &54684--59573   &54303--59182 \\ 
J1718$-$3825$^{e}$  &17:18:13.565(4)  &$-$38:25:18.06(15) &0.074675  &13.1742  &1.00   &89.8  &247.46(6)    &54741--59491   &54303--59238 \\ 
\hline 
    \hline    
\end{tabular}
\begin{tablenotes}
\item[ ] \textit{Note}. References for parameters of these pulsars: 
\item[ ] $^a$ \cite{KeithJKW2008};  
\item[ ] $^b$ \cite{D'AmicoKMC2001}; 
\item[ ] $^c$ \cite{KramerBML2003};  
\item[ ] $^d$ \cite{JohnstonSMK1992}; 
\item[ ] $^e$ \cite{ManchesterLCB2001}.
\end{tablenotes} \vspace{-0.2cm}
\end{table*}

\section{Introduction} \label{sec:intro}
In pulsar timing observations, the phenomenon that the spin frequency jumps in a short time is called a glitch.
Glitch is a rare phenomenon, currently detected in only about 6$\%$ of pulsars \citep{1996MNRAS.282..677S,2022RPPh...85l6901A}.
Studies have found that most glitches occur in relatively young isolated pulsars with ages spanning between $10^{3}$ to $10^{5}$ yr \citep{YuMHJ2013,BasuAKL2022}, the most famous being the Vela (PSR J0835$-$4510; \citealt{RadhakrishnanM1969,ReichleyD1969,2024A&A...689A.191Z,grover2024}) and Crab pulsars (PSR J0534+2200; \citealt{EspinozaSK2011}).
It is worth noting that glitch is not exclusive to young pulsars, but also occur in old pulsars. 
For instance, millisecond pulsars such as PSRs J0613$-$0200 and J1824$-$2452 (B1821$-$24) have exhibited minor glitches with fractional glitch sizes around $\sim10^{-12}$ \citep{CognardB2004, McKeeSLC2016}.
Moreover, the intriguing occurrence of glitches extends to magnetars including 1E 2259+586 \citep{SasmazAG2014}, as well as to accretion-powered pulsar SXP 1062 \citep{SerimSDS2017} and binary pulsars (PSR J1915+1606; \citealt{WeisbergNT2010}).

Young gamma-ray pulsars, characterised by showing high spin-down
rates active in the gamma-ray band of emission, are good target sources for searching and studying glitches \citep{SokolovaP2022}.
More than 150 young gamma-ray pulsars have been reported so far, with more than 80 among them detected also in
radio emissions \citep{SmithAAB2023}.
Investigating glitches in gamma-ray pulsars holds the potential to not only provide valuable insights into the internal structure of neutron stars but also to offer a window into the dynamic evolution of the magnetosphere, see e.g., \citet{KouYWY2018}.

Currently, there has been notable progress in research focused on glitches in gamma-ray pulsars.
PSR J2021+4026 is the only glitching pulsar in which a gamma-ray flux jump has been observed \citep{AllafortBBB2013,TakataWLH2020}.
The studies by \citet{NgTC2016} and \citet{ZhaoNLT2017} indicated that the gamma-ray flux jumps and the corresponding substantial shifts in the pulse profile of PSR J2021+4026 might have a plausible association with glitches.
In a similar manner, \citet{LinWHT2021} reported the detection of a new glitch in PSR J1420$-$6048 using \textit{Fermi}-LAT 11 yr timing data and found that the gamma-ray pulse profile changed significantly before and after glitches 4 and 8.
\citet{2022MNRAS.511..425G} identified 20 glitches in 4 gamma-ray pulsars, and made a relatively accurate prediction of the waiting time for the next glitch through the vortex creep model.

In this paper, we study the glitches of five gamma-ray pulsars by combining the timing observation data of the \textit{Fermi}-LAT and Parkes telescopes over the past 14 yr.
We have identified 12 glitches in total and detected a new small glitch in PSR J1718$-$3825.
Besides, we update the epoch and other parameters of the previously reported 11 glitches.

The structure of this paper is as follows: 
In Section 2, we briefly introduce the timing observations from \textit{Fermi}-LAT and Parkes 64 m radio telescope and give an overview of the glitch data processing procedure.
One superfluid model, namely the vortex creep model, that is used in glitch fitting is introduced in Section 3, following which we relate the glitch observables to the dynamical and structural properties of the neutron star. 
The results and underlying physical process of pulsar glitches studied in the paper are detailed in Section 4. We then discuss our results more extensively in Section 5 and finally summarize our findings in Section 6.
 
\begin{table*} 
\caption{Pre- and post-glitch timing solutions obtained using {\scriptsize \text{TEMPO}2}, with a parameter uncertainty of 1$\sigma$.
}\label{Tab:F0F1-works}
    \setlength{\tabcolsep}{8.5pt}      
        \renewcommand{\arraystretch}{1.25}
\begin{tabular}{cccccccccc}
  \hline \hline
PSR & Int. & Epoch & $\nu$ & $\dot{\nu}$         & $\ddot{\nu}$        & $N_{\rm ToA}$ & $T$    & Data span & RMS \\
 &{ }   & (MJD)  & (Hz)  & (10$^{-14}$s$^{-2}$) & (10$^{-24}$s$^{-3}$) & { }       & (yr) & (MJD)   & ($\upmu$s) \\      
    \hline     
J1028$-$5819 &0--1  &56292  &10.94024439333(3) &$-$192.67239(3)   &7.72(1)  &380  &8.8  & 54687--57898   &1536 \\
             &1--    &58858  &{10.9399877351(3)}  &$-$193.3893(6)    &31.0(5)  &221  &5.2  & 57909 -- 59809   &8827 \\
J1420$-$6048 &3--4  &54542  &14.6625943767(3)  &$-$1782.52(1)      &1130(18) &16   &0.9  & 54303--54634   &2  \\  
            &4--5   &55279  &14.661471259(2)  &$-$1781.60(3)      &1549(12) &58   &2.0  & 54652--55397   &5256   \\  
            &5--6   &55841  &14.660623935(1)  &$-$1785.510(4)     &1003(7)  &53   &2.3  & 55408--56256   &3950   \\  
            &6--7   &56704  &14.659321169(2)  &$-$1785.782(9)     &714(14)  &51   &2.6  & 56256--57212   &9932   \\  
            &7--8   &57890  &14.657509511(3)  &$-$1783.284(6)     &765(7)   &62   &3.6  & 57228--58555   &17187  \\  
            &8--     &59030  &14.655774598(1)  &$-$1784.302(5)     &983(7)   &38   &2.5  & 58557--59453   &3805  \\             
J1709$-$4429 &3--4  &54496  &9.7566104500(1)  &$-$883.9007(7)    &330(3)    &21   &1.1  & 54303--54690   &30 \\ 
            &4--5   &55516  &9.7558562250(5)  &$-$885.766(1)     &239.2(9)  &435  &4.5  & 54693--56337   &17667  \\ 
            &5--6   &57256  &9.7545531968(5)  &$-$885.721(1)     &197.8(8)  &493  &5.0  & 56340--58173   &21964  \\ 
            &6--     &{58874}  &{9.7533385441(9)}  &{$-$886.015(2)}     &{229(2)}    &249  &3.8  & 58173--59582   &19600  \\         
J1718$-$3825 &0--1  &54605  &{13.3915736271(3)} &{$-$236.412(1)}     &{28(3)}       &24   &2.2  & 54303--54920   &513 \\    
            &1--2   &56466  &{13.39119382383(4)} &{$-$236.05539(7)}     &{20.75(3)}  &106  &8.2  & 54997--57940   &104\\  
            &2--3   &58524  &{13.3907744744(1)} &{$-$235.7702(3)}     &{17.2(3)}       &45   &3.2  & 57964--59113   &533\\ 
            &3--     &59301  &13.3906162607(6) &$-$235.65(1)   &$-$      &8    &1.1  & 59128 -- 59491   &988\\ 
J1509$-$5850 & --    &56906  &11.2452960325(2)  &$-$115.9029(5)    &$-$       &172  &14.3 & 54303--59510   &34250\\        
\hline    \hline
\end{tabular} 
\end{table*}

\section{Observation and analysis}
\label{sect:Orb}

In this study, we focus on the multi-band research of pulsar glitches. The selected pulsars for this study must meet two criteria simultaneously: first, they exhibit emissions in both radio and gamma-ray wavelengths; second, they have more than 10 yr of high-quality timing data available through long-term monitoring by the \textit{Fermi}-LAT and the Parkes 64 m radio telescope.
Based on these criteria, we selected five young gamma-ray pulsars with spin evolution similar to Vela pulsar for this study, namely PSRs J1028$-$5819, J1420$-$6048, J1509$-$5850, J1709$-$4429 and J1718$-$3825, with timing data spanning from 2007 to 2022.
Table \ref{Tab:publ-works} presents the detailed parameters of pulsars.
The first three columns are the pulsar name (J2000) and the position [right ascension (RA) and declination (DEC)] in the equatorial coordinate system.
The fourth and fifth columns are the pulsar spin period ($P$) and its first time derivative ($\dot{P}$).
The sixth and seventh columns are the inferred surface dipole magnetic field strength at the stellar equator ($B_{\rm s} = 3.2\times 10^{19} \sqrt{P\dot{P}}$ G) and the characteristic age ($\tau_{\rm c}=P/2\dot{P}$) of pulsars.
The last three columns are the pulsar's dispersion measure (DM) and the timing observation data time span of the \textit{Fermi}-LAT and the Parkes telescopes for the corresponding sources.

The \textit{Fermi} satellite was launched in 2008, and the LAT is the main instrument, which is a great field of view (2.4 sr) gamma-ray telescope, and it scans the entire sky every 3 h \citep{AtwoodAAA2009,AbdoAAA2010}.
\textit{Fermi}-LAT can receive gamma-ray photons in the energy range of 20 MeV--1 TeV, with an effective receiving area of 9000 $\rm cm^{2}$ at 2--500 GeV \citep{AbdollahiAAA2020}.
Its photon collecting time accuracy is higher than 0.3 $\upmu$s, so that millisecond pulsars can be easily identified.
In the last 16 yr, \textit{Fermi}-LAT has detected 340 gamma-ray pulsars and candidates \citep{SmithAAB2023}.
This provides us with a valuable data to analyse gamma-ray pulsar timing. 
We used \textit{Fermi} Science Tool (v11r5p3)\footnote{\href{https://fermi.gsfc.nasa.gov/ssc/data/analysis/scitools/}{https://fermi.gsfc.nasa.gov/ssc/data/analysis/scitools/}} for timing analysis. 
The events were constructed using gtselect with an angular distance less than $0.5^{\circ}$, zenith angle less than $105^{\circ}$, and an energy range of 100--10000 MeV \citep{RayKPA2011}.
Then, the arrival time of each event was corrected to the Solar system barycentre (SSB) using gtbary to obtain the pulse times-of-arrival (ToAs). 
Each ToA was accumulated from several days of exposure with an accumulated time of 10, 50, 60, 4, and 90 d for PSRs J1028$-$5819, J1420$-$6048, J1509$-$5850, J1709$-$4429, and J1718$-$3825, respectively. 
In the data from 2008 to 2022, a total of 505, 96, 81, 1084, and 48 ToAs were obtained for these pulsars, respectively.
The detailed procedure can be found in \cite{ge2019}.

We also collected the timing observation data of these five gamma-ray pulsars from 2007 to 2020 on the Parkes pulsar data archive\footnote{\href{https://data.csiro.au/domain/atnf}{https://data.csiro.au/domain/atnf}}. 
\cite{ManchesterHBC2013} has introduced the timing observation of Parkes 64 m radio telescope in detail. 
Briefly, Parkes observation was carried out at 20 cm band, mainly using the multi-beam receiver, with a centre frequency of 1369 MHz and bandwidth of 256 MHz \citep{StaveleyWBD1996,HobbsMDJ2020}.
The Parkes Digital Filter Bank systems (including PDFB1, PDFB2, PDFB3, and PDFB4) are used for data recording. The raw data consists of 1024 frequency channels, each with a bandwidth of 0.25 MHz. The integration times were 2 to 15 mins, with a sub-integration time of 30 s and the observation intervals of 2 to 4 weeks for each pulsar.

With regarding to the radio observations, the {\footnotesize \text{PSRCHIVE}} package was used to process the raw data. After eliminating radio frequency interference (RFI), data are de-dispersed in frequency and summed in frequency and time to obtain an integrated pulse profile for each observing session \citep{Hotan2004,StratenDO2012}. The \texttt{PAT} tool of {\footnotesize \text{PSRCHIVE}} was then used to cross-correlate each mean profile with the standard profile, leading to produce the time-of-arrivals (ToAs; \citealt{Taylor1992}). The standard profile is formed by aligning the pulse phase and summing all the observations. Next, the Jet Propulsion Laboratory planetary ephemeris DE440 \citep{ParkFWB2021} and the Barycentric Coordinate Time (TCB) were used to convert these local ToAs to (Solar System Barycenter) SSB, thereby eliminating the effect of the Earth motion on the ToAs.

When combining the gamma-ray and radio ToAs, to ensure the phase alignment of the data from two telescopes,
the package {\footnotesize \text{TEMPO}}2 \citep{HobbsEM2006,EdwardsHM2006} has been used to fit the phase offset of the data of the Parkes telescope and \textit{Fermi}-LAT. The phase offset value obtained from this fitting will be appended to the ephemeris to achieve phase alignment of the timing data of the two telescopes.
We also introduced two parameters, EFAC and EQUAD, to correct the uncertainty of each ToA acquired from different telescopes \citep{WangHCS2015}.
Among them, EFAC is a free parameter that adjusts unaccounted instrumental errors, and EQUAD accounts for additional time-independent white noise processes \citep{LentatiAHF2014}.
The values of these two parameters were determined by the \texttt{EFACEQUAD} plug-in of {\footnotesize \text{TEMPO}}2, and this method did not involve Bayesian techniques.
After correction, the relationship between the initial uncertainty ($\delta$) of ToAs and its new value (${\delta}_{\rm s}$) was defined as ${\delta}_{\rm s}^{\rm 2} = {\rm EFAC}^{2} \times(\delta^{2}+{\rm EQUAD}^{2})$.
After adding the error correction of ToA, the reduced chi-square values of each fitting region are distributed between 0.95--1.05.

The pulsar position is one important factor that probably affects the precision of the glitch parameters, but for some sources, the position obtained from Pulsar Catalogue ({\footnotesize \text{PSRCAT;  \citealt{ManchesterHTH2005}}})\footnote{\href{https://www.atnf.csiro.au/research/pulsar/psrcat/}{https://www.atnf.csiro.au/research/pulsar/psrcat/}} is not accurate enough. 
Although the accurate positions of most bright pulsars can be determined by VLBI (Very Long Baseline Interferometry), the positions of most relatively weak pulsars could be constrained by pulsar timing. For these pulsars, we used the "Cholesky" method to minimize the influence of timing noise on the determination of the pulsar position \citep{ColesHCM2011}.
To do this, we selected a sufficiently long length of data that not include any glitch, invoked the {\footnotesize \text{TEMPO}}2 plug-in \texttt{SPECTRALMODEL} to fit the red noise model (power law model of red noise: $P(f) = {A}{[1+(f/f_{\rm c})^{2}]^{(-\alpha/2)}}$, where $A$, $\alpha$, and $f_{\rm c}$ are amplitude, spectral index, and corner frequency, respectively).
Then, by analyzing this red noise model and applying a global least-squares fitting procedure, we achieved precise fitting of the pulsar position parameters.
For example, the position parameters of PSR J1028$-$5832 has been re-determined, updating the RA to 10:28:27.9(1) and the DEC to $-$58:19:06.21(6). These updated values are included in Table 1. Compared to the cataloged position [RA 10:28:28.0(1) and DEC $-$58:19:05.2(15)] provided by PSRCAT, the RA remains nearly unchanged, while the DEC shows a significant shift of approximately $1^{\arcsec}$. Additionally, the uncertainty in the DEC has been substantially reduced.
For the other four pulsars, as their new positions are consistent with previously published data, we used the previously reported positions in the timing analysis.
Note that the red noise model is exclusively utilized for deriving the pulsar position parameters and is not applied to other data analysis processes.

After completing the above steps, we employed {\footnotesize \text{TEMPO}}2 to fit ToAs to obtain the best rotation phase model of pulsars.  
By subtracting the actual observed ToAs from the ToAs predicted by the model, timing residuals were obtained.

According to the pulsar timing model, the evolution of pulse phase with time, $\phi(t)$, can be expressed as \citep{EdwardsHM2006}:
\begin{equation}
\label{equ:1} 
\phi(t) = \phi_{\rm 0} + \nu(t - t_{\rm 0}) + \frac{\dot{\nu}}{2!} (t - t_{\rm 0})^{2} + \frac{\ddot{\nu}}{3!}(t - t_{\rm 0})^{3} \ ,
\end{equation}
where $t_{\rm 0}$ represent certain initial epoch, and $\phi_{\rm 0}$ is the initial pulse phase at $t = t_{\rm 0}$. Here, $\nu$ is the spin frequency of pulsar,  $\dot{\nu}$ and $\ddot{\nu}$ represent the first and second derivatives of the frequency, respectively.

We can use {\footnotesize \text{TEMPO}}2 to visualize timing residuals.
If the timing model is correct, the timing residuals should be randomly distributed around zero. If, upon visual inspection, the residuals appear discontinuous or show significant deflection, it may indicate the occurrence of a glitch event in the pulsar.

Assuming a glitch event occurs at epoch $t_{\rm g}$, it may result in a change in the pulse phase is $\Delta\phi$, a permanent increase in the spin frequency, $\Delta \nu_{\rm p}$, and a permanent increase in its first derivative $\Delta\dot{\nu}_{\rm p}$.
It should be noted that the post-glitch recovery process typically involves one or more exponential decay processes.
If the frequency change during the exponential recovery process of the $i$-th phase is $\Delta\nu_{\rm d_{i}}$, and the corresponding timescale is $\tau_{\rm d_{i}}$, we can describe the change of post-glitch pulse phase according to the above-mentioned relevant glitch parameters as \citep{EdwardsHM2006, YuMHJ2013}:
\footnote{Although $\ddot\nu$ may vary following a glitch, this variation is likely contaminated by the post-glitch recovery process, $\Delta\ddot\nu$ was excluded from Eq. (\ref{equ:2}).}
\begin{equation}
\begin{split}
\label{equ:2} 
\phi_{\rm g} = &\Delta\phi+ \Delta\nu_{\rm p}(t - t_{\rm g}) +   \frac{1} {2} \Delta\dot{\nu}_{\rm p} (t - t_{\rm g})^{2} \\
&+ \sum_{\rm i} {\Delta\nu_{\rm d_{i}}\tau_{\rm d_{i}} [1-e^{-(t - t_{\rm g})/\tau_{\rm d_{i}}}] }\ . 
\end{split}
\end{equation}
According to the above equation, the total fractional glitch size and its first derivative are respectively represented as:
\begin{equation}
\frac{\Delta{\nu}}{{\nu}} = \frac{\Delta{\nu}_{\rm p} + \sum_{\rm  i}\Delta\nu_{\rm d_{i}}} {\nu} \ ,
\end{equation}
and
\begin{equation}
 \frac{\Delta\dot{\nu}} {\dot{\nu}} = \frac{\Delta\dot{\nu}_{\rm p} - \sum_{\rm i} {\Delta\nu_{\rm d_{i}} / \tau_{\rm d_{i}}}} {\dot{\nu}} \ .
\end{equation}
Finally, the parameter $Q$ that quantifies the fraction of glitch recovery
can be defined as: 
\begin{equation}
 Q=\frac{ \sum_{\rm i} {\Delta\nu_{\rm d_{i}}} } {\sum_{\rm i} {\Delta\nu_{\rm d_{i}}} + \Delta\nu_{\rm p}} \ .   
\end{equation}

\begin{table*} 
\centering
\caption{Glitch parameters obtained through fitting methods based on Eqs. (\ref{equ:1}) and (\ref{equ:2}). See text for details.
 }\label{Tab:glitch-works}
    \renewcommand{\arraystretch}{1.4} 
     \setlength{\tabcolsep}{10pt}      
\begin{tabular}{ccccccccccc}
  \hline  \hline
PSR  & Gl. No. & Epoch & $\Delta \nu/\nu$ & $\Delta \dot{\nu}/\dot{\nu}$  & NEW? & $Q$ & $\tau_{\rm d}$ & RMS & Data span \\
& {  } & (MJD)  & (10$^{-9}$) & (10$^{-3}$) & (Y/P) & {  }  & (d) & ($\upmu$s) & (MJD)  \\       
    \hline  
J1028$-$5819  & 1  & 57904(6)  & 2284.0(9)   &  20(2)  & P &  0.0060(2) &  62(7) & 636 & 57358--58243 \\
J1420$-$6048  & 4  & 54653(19) &  937.6(9)  &  6.4(3)   & P &  0.0119(6) &  53(7)  & 3  & 54310--54949 \\
             & 5  & 55400(9)  &  1367(15)   &  5.3(8)   & P &  0.016(8)  &  187(82) & 991      & 55040--55804  \\
             & 6  & 56256(1)  &  1967(2)   &  8.6(9)    & P &  0.0099(7)  &  44(8) & 818      & 55956--56536  \\
             & 7  & 57216(12) &  1206(1)    &  4.9(2)     & P &  0.0183(6)  &  102(7) & 943    & 56800--58110  \\
             & 8  & 58555(2)  &  1480(1)   &  6.0(1)     & P &  0.0131(5) &  100(6)  & 434       & 58141--59002   \\
J1709$-$4429  & 4  & 54691(2)  &  2777(4)   &  64(15)    & P  &  0.0104(5) &  72(4)    & 387       & 55415--55167  \\
             &    &           &           &            &   &  0.006(1)  &  4.2(6)    &            &                 \\
             & 5  & 56339(2) &  2964(4)   &  9.7(8)     & P &  0.0095(8) &  61(6) & 747     & 56158--56603   \\          
             & 6  & 58175(2)  &  2438(3)   &  10(1)     & P &  0.0087(7) &  47(8)  & 1189     & 57918--58401                 \\

J1718$-$3825 & 1  & 54952(44) &  1.98(6)    &  $-$0.10(3)  & P & --    & --    & 50    & 54289--55860  \\
             & 2  & 57950(14) &  7.37(9)    &  0.21(6)     & P & --       & --   & 493   & 57160--58515  \\
             & 3  & 59121(8)  &  1.9(2)     &  $-$0.12(9)  & Y & --      & --    & 596   & 58510--59527  \\
\hline    \hline    
\end{tabular}
\end{table*}

\begin{table*}
\caption{Vortex creep model fit results for 9 large glitches in three gamma-ray pulsars analysed in this work:
The characteristic relaxation time of exponential decay $\tau_{\rm creep,exp}$ from the linear response regime in vortex creep model can be confronted with that of the {\scriptsize \text{TEMPO}}2 fitting $\tau_{\rm d}$ listed in Table \ref{Tab:glitch-works}. 
$\tau_{\rm nl}$ ($\tau_{\rm s}$ in the case of single non-linear regime) is the non-linear creep relaxation time. 
Post-glitch recovery is expected to end when the relaxations of the non-linear creep regions are over, and glitch induced changes in the spin-down rate recover completely after a waiting time $t_0$. Thus the theoretical expectation $t_{\rm th}=\mbox{max}(t_{0},t_{\rm single})$, i.e., the maximum of Eq. (\ref{t0}) and Eq. (\ref{tsingle}) gives a prediction for the inter-glitch time. The observed inter-glitch time $t_{\mathrm{obs}}$ is given for comparison with the theoretical prediction by the vortex creep model $t_{\rm th}$. 
}\label{creepmodelfit}
     \setlength{\tabcolsep}{13pt}      
\begin{tabular}{lcccccccc}
\hline\hline
PSR & Gl. No. & $\tau_{\rm creep,exp}$ & $\tau_{\rm nl}$ & $t_{0}$ &  $\tau_{\rm s}$ &  $t_{\rm s}$ & $t_{\rm th}$ & $t_{\rm obs}$  \\
&   & (d)  & (d) & (d) & (d) & (d)  & (d) & (d)  \\ 
\hline
J1028$-$5819 & 1  & 96(12) &71 & 4890(269) & $-$ & $-$ & 4890(269) & $-$ \\
J1420$-$6048 & 4 & 41(11) & $-$ & $-$ & 116(10) & 349(8) & 698(32) & 747(28) \\
             & 5  & $-$  & 109(23) & 125(42)  & 109(6) & 638(9) & 964(19) & 856(10)\\
             & 6 & 84(11)  & 20(12) & 562(13)  & 98(11) & 829(6) & 1124(32) & 960(13) \\
             & 7 & 109(32) & 779(554) & 938(425) & $-$ & $-$  & 938(425) & 1339(14) \\
             & 8 & 86(39) & 19 & 823(16) & 26(11) & 623(11) & 823(16) & $-$ \\
J1709$-$4429 & 4 & 86(9) & 3.4 & 1680(16) & $-$ & $-$ & 1680(16) & 1648(4) \\
             & 5 & 95(19)  & 28 & 1861(25) & $-$ & $-$ & 1861(25) & 1836(4) \\
             & 6 & 93(9) & 3.2 & 1850(26) & $-$ & $-$  & 1850(26) & $-$ \\
\hline\hline
\end{tabular}
\end{table*}

\section {Glitch trigger and post-glitch relaxation in the vortex creep model}
\label{sec:model}
The observations of pulsar glitches provided the first evidence of superfluidity in neutron-star interiors~\citep{1969Natur.224..673B}. Previously,  our works focused on the amplitude of the glitch and explored whether modern equation of state (EOS) of pulsars can reconcile with the available data~\citep{2015ChPhL..32g9701L,2016ApJS..223...16L}, mainly in the framework of the two-component model~\citep{1969Natur.224..673B,1975Natur.256...25A}, i.e., the (crustal) superfluid component and the other is the charged component consisting of all matter in the normal phase and the proton-neutron fluid conglomerate coupled to each other via electrons scattering off vortex lines and strong magnetic field. With various post-glitch recovery data 
available~\citep{2020ApJ...896...55G,2022MNRAS.511..425G,LiuYGY2024}, we here adopt the vortex creep model and confront it with both the observed amplitudes and post-glitch decays~\citep{2021ApJ...923..108S}, in our endeavor to search for a firmly grounded paradigm of glitch phenomenon. 

The vortex creep model \citep{AlparAS1984,Alpar1989} aimed at explaining both the glitch triggering and the post-glitch relaxation.
In the theory, there exist a number of distinct superfluid regions  with different pinning energies in the inner crust. The neutron superfluid vortices are pinned to nuclei or interstitial positions, developing a lag $\omega=\Omega_\mathrm{s}-\Omega_\mathrm{c}$ between the pinned neutron superfluid and the crustal component, which spins down under the external pulsar braking torque. Here $\Omega_\mathrm{s}$ and $\Omega_{\rm c}$ are the superfluid and the crustal angular velocities, respectively. The radially outward motion of vortices caused by thermal excitation overcoming pinning barriers is termed vortex creep. The 
continuous vortex creep results in angular momentum transfer from the crustal superfluid to the crust and the coupling between them, a steady state $\omega_{\infty}$ can reached when the crustal superfluid and crust share the same spin-down rate $\dot{\Omega}_\mathrm{s}=\dot{\Omega}_{\rm c}=\dot{\Omega}_{\infty}$. In the steady state, the star's spin-down appears as if an external torque $N_{\rm{ext}}$ acts on the star's total moment of inertia $I$, $N_{\rm{ext}}=I\dot{\Omega}_{\infty}$.

A glitch occurs if a local fluctuation raise $\omega$ to above $\omega_\mathrm{cr}$, here $\omega_\mathrm{cr}$ is a critical lag where the Magnus force exceeds the pinning force and it depends on the region where the vortices are located. When a glitch occurs, a number of vortices unpin suddenly from pinning sites and move outward; at the same time, the steady state is broken down with a perturbation $\delta\omega$ deviated from $\omega_{\infty}$. Due to the conservation of angular momentum, the angular velocity jump $\Delta\Omega_\mathrm{c}$ for the observed crust is
\begin{equation}\label{conserJ}
    I_\mathrm{c}\Delta\Omega_c=\sum_iI_{\rm i}\delta\Omega_{\rm i},
\end{equation}
where $I_{\rm i}$ are the moment of inertia of distinct superfluid regions and $\delta\Omega_{\rm i}$ are the corresponding decrease in the angular velocity. The derivation $\delta\omega$ in a superfluid region can be expressed by $\delta\omega_{\rm i}=\Delta\Omega_\mathrm{c}+\delta\Omega_{\rm i}$. After a glitch, the star returns to a new steady state, towards original pre-glitch state, as a result of the recovery of vortex creep. Depending on the temperature and the pinning properties, the vortex creep can have a linear or non-linear response to the perturbation $\delta\omega$ in the different superfluid regions \citep{Alpar1989}.

In the linear response regime, approach to the steady state has a linear dependence on the initial perturbation $\delta\omega(0)$. The post-glitch recovery in this regime can be written by
\begin{equation}\label{linear_response}
    \Delta\dot{\Omega}_{\mathrm{c,i}}(t)=-\frac{I_{\rm i}}I\frac{\delta\omega_{\rm i}(0)\rm{e}^{-t/\tau_{\mathrm{l,i}}}}{\tau_{\mathrm{l,i}}},
\end{equation}
where $\tau_{\mathrm{l,i}}$ is the relaxation time of this regime and $\delta\omega_{\rm i}(0)$ is the value of $\delta\omega_{\rm i}$ at the time when a glitch occurs. The relaxation behaviour is exponential decay with the characteristic relaxation time
\begin{equation}\label{linear_relaxtime}
    \tau_\mathrm{l}=\frac{kT}{E_\mathrm{p}}\frac{\omega_\mathrm{cr}r}{4\Omega_{\rm c} v_0}\exp(\frac{E_\mathrm{p}}{kT}),
\end{equation}
where $kT$ and $E_\mathrm{p}$ are the internal temperature and the pinning energy in this
regime; $r$ is the distance of this regime from the spin axis; $v_0 \approx 10^5$--$10^7$ cm s$^{-1}$ is the typical velocity of microscopic vortex motion \citep{erbil16,erbil2023}. 
In the outer core, vortex creep against flux tubes gives a similar post-glitch exponential relaxation with the relaxation time \citep{erbil14},
\begin{align}\label{tautor}
    \tau_{\rm tor}\simeq &60\left(\frac{\vert\dot{\Omega}_\mathrm{c}\vert}{10^{-10}\mbox{
    \radss}}\right)^{-1}\left(\frac{T}{10^{8}\mbox{
    K}}\right)\left(\frac{R}{10^{6}\mbox{ cm}}\right)^{-1}
    x_{\rm p}^{1/2}\times\nonumber\\&\left(\frac{m_{\rm p}^*}{m_{\rm p}}\right)^{-1/2}\left(\frac{\rho}{10^{14}\mbox{\gcc}}\right)^{-1/2}\left(\frac{B_{\phi}}{10^{14}\mbox{ G}}\right)^{1/2}\mbox{d}\ ,
\end{align}
where $\dot{\Omega}_\mathrm{c}$ is the spin-down rate, $R$ is the radius of the outer core, $x_{\rm p}$ is the proton fraction in the neutron star core, $\rho$ is the matter density, $ m_{\rm p}^*(m_{\rm p})$ is the effective (bare) mass of protons, and $B_{\phi}$ is the toroidal component of the magnetic field strength. A linear regime occurs when the temperature is high enough compared to the pinning energy, hence we have $\omega_{\infty}\ll\omega_\mathrm{cr}$ and $\delta\omega_{\rm i}(0)\approx\Delta\Omega_\mathrm{c}$.

In the non-linear response regime, at the other extreme, approach to the steady state depends non-linearly on the initial perturbation $\delta\omega(0)$. For a single crustal superfluid region, $k$, in the non-linear regime, the post-glitch response is
\begin{equation}\label{nonlinear_response}
    \frac{\Delta\dot{\Omega}_{\mathrm{c},k}(t)}{\dot{\Omega}_\mathrm{c}}=-\frac{I_{\rm s,k}}I\left\{1-\frac1{1+[\rm{e}^{(t_{\mathrm{0},k}/\tau_{\mathrm{n},k})}-1]\rm{e}^{(-t/\tau_{\mathrm{n},k})}}\right\}\ ,
\end{equation}
where
\begin{equation}\label{t0}
t_{\mathrm{0,k}}=\delta\omega_{\rm k}(0)/|\dot{\Omega}|_{\infty}\ ,
\end{equation}
and
\begin{equation}\label{nonlinear_relaxtime}
    \tau_\mathrm{n}=\frac{kT}{E_\mathrm{p}}\frac{\omega_\mathrm{cr}}{|\dot{\Omega}|_{\infty}}\ .
\end{equation}
Because $\dot{\Omega}_{\rm s}=\dot{\Omega}_{\rm c}=\dot{\Omega}_{\infty}$ in the steady state and the magnitude of $\Delta\dot{\Omega}_{\rm c}$ is small compared to $\dot{\Omega}_{\rm c}$, $\Delta\dot{\Omega}_\mathrm{c}/\dot{\Omega}_\mathrm{c}\sim10^{-2}$, we use observed value of $|\dot{\Omega}_{\rm c}|$ to represent $|\dot{\Omega}|_\infty$. In the non-linear regime, the interior temperature is sufficiently low compared to the pinning energy, hence we have $\omega_{\infty}\lesssim\omega_\mathrm{cr}$ and $\delta\omega_{\rm k}(0)\approx\delta\Omega_{\rm s,k}$. Eq. (\ref{nonlinear_response}) has the form of making a ``shoulder'' such that at about $t_\mathrm{0,k}$ after the glitch, the spin-down rate decreases rapidly by $I_{\rm s,k}/I$ in a time
$\sim3\tau_{\mathrm{n,k}}$ quasi exponentially. This response resembles the behaviour of the Fermi distribution function of statistical mechanics for non-zero temperatures \citep{AlparAS1984}. Single superfluid region’s recovery is
thus completed on a timescale,
\begin{equation}\label{tsingle}
    t_{\rm{single,k},} = t_{\rm 0,k}+3\tau_{\mathrm{n,k}}.
\end{equation}

In several neighbouring single non-linear regime regions, we can integrate explicitly the contributions from these single non-linear regime regions by making the assumption that the post-glitch superfluid angular velocity decreases linearly in $r$ from $\delta\Omega_\mathrm{A}$ over the region \citep{AlparAS1984}, then the post-glitch response is \citep{alpar96}:
\begin{equation}\label{integ_nonlinear_response}
    \frac{\Delta\dot{\Omega}_\mathrm{c,A}(t)}{\dot{\Omega}_\mathrm{c}}=\frac{I_\mathrm{A}}{I}\left\{1-\frac{1-(\tau_\mathrm{n,A}/t_\mathrm{0,A})\ln\left[1+(\mathrm{e}^{t_\mathrm{0,A}/\tau_\mathrm{n,A}}-1)\mathrm{e}^{-\frac{t}{\tau_\mathrm{n,A}}}\right]}{1-\mathrm{e}^{-\frac{t}{\tau_\mathrm{n,A}}}}\right\},
\end{equation}
Here, the total moment of inertia of the non-linear creep regions affected by vortex unpinning events is denoted by $I_\mathrm{A}$. \citet{erbil20} obtained the following estimate for $\tau_{\rm{n}}$ by incorporating density dependence of the pinning energy throughout the crustal superfluid and temperature evolution of the neutron star,
\begin{equation}
\tau_{\rm{n}}\cong(30-1420)\left(\frac{\dot\nu}{10^{-11} \mbox{\,\rm s$^{-2}$}}\right)^{-1}\left(\frac{\tau_{\rm c}}{10^{4} {\,\rm yr}}\right)^{-1/6}\mbox{d},
\label{nonlinear_relaxtime_erbil}
\end{equation}
where $\tau_{\rm c}=\nu/(2\dot\nu)$ is the spin-down timescale (characteristic age) of the pulsar. In the limit of $t_\mathrm{0,A}\gg\tau_\mathrm{n,A}$, Eq. (\ref{integ_nonlinear_response}) reduces to
\begin{equation}\label{reduced_integ_nonlinear_response}
    \frac{\Delta\dot{\Omega}_\mathrm{c,A}(t)}{\dot{\Omega}_\mathrm{c}}=\frac{I_\mathrm{A}}{I}\left(1-\frac{t}{t_\mathrm{0,A}}\right),
\end{equation}
with a constant, anomalously large $\Ddot{\Omega}_\mathrm{c}=(I_{\rm A}/I)(|\dot{\Omega}_{\rm c}|/t_{\rm 0,A})$ \citep{AlparBA2006}. The post-glitch recovery is expected to come to end when the relaxations of the non-linear creep regions are over. So the theoretical expectation $t_{\rm th}=\mbox{max}(t_\mathrm{0,A},t_{\rm single})$, i.e. maximum of Eq.~(\ref{t0}) and Eq.~(\ref{tsingle}) gives a prediction for the inter-glitch time. In practice, we usually choose a summation of several post-glitch recoveries of linear and non-linear regions to fit the post-glitch observations, e.g., a linear region with $I_{\rm{exp}}=I_1$ and $\tau_{\rm{creep,exp}}=\tau_{\rm l,1}$ in Eq. (\ref{linear_response}), a single non-linear region with $I_{\rm{s}}=I_{\rm s,1}$, $t_{\rm{s}}=t_{0,1}$ and $\tau_{\rm{s}}=\tau_{\rm n,1}$ in Eq. (\ref{nonlinear_response}), and a continuous nonlinear regions with $I_{\rm{A}}$, $t_0=t_{0,\rm A}$, and $\tau_{\rm{nl}}=\tau_{\rm{n,A}}$ in Eq. (\ref{integ_nonlinear_response}). $\Delta\dot{\Omega}_\mathrm{c}/\dot{\Omega}_\mathrm{c}$ is related to $\Delta\dot{\nu}/\dot{\nu}$ by $\Omega=2\pi\nu$.
\begin{table}
\caption{Moment of inertia of distinct superfluid regions for 9 large glitches in three gamma-ray pulsars analysed in this work: $I_{\rm A}$ is the total moment of inertia of the non-linear creep regions affected by vortex unpinning events, i.e., $I_{\rm A}/I$ is the fractional moment of inertia of the superfluid regions that gave rise to the glitch by collective unpinning event; the vortex then move to other trapped region by crossing vortex-free region B with moment of inertia $I_{\rm B}$, i.e., $I_{\rm B}/I$ is the fractional moment of inertia of the superfluid regions within which unpinned vortices moved;
$I_{\rm exp}/I$ is fractional moment of inertia of the exponentially decaying superfluid region.
$I_{\rm cs}/I$ is the fractional moment of inertia of the crustal superfluid that involved in the corresponding glitch event, which corrected for crustal entrainment effect average enhancement factor $\langle m_{\rm n}^{*}/m_{\rm n}\rangle=5$ \citep{delsate16}.
}\label{moi}
    \setlength{\tabcolsep}{1.5pt}      
        \renewcommand{\arraystretch}{1.15}
\begin{tabular}{lccccccc}
\hline\hline
PSR & Gl. No.& $I_{\rm exp}/I$ & $I_{\rm A}/I$ & $I_{\rm s}/I$ &  $I_{\rm B}/I$ &   $I_{\rm cs}/I$ \\
&  & ($10^{-3}$)  & ($10^{-3}$)  & ($10^{-3}$) & ($\%$)  &  ($\%$) \\ 
\hline
J1028$-$5819 & 1 & 7.92(1.63)  & 5.25(8)   & $-$ & 2.83(16)  & 3.35(16)\\
J1420$-$6048 & 4 & 15.34(6.41)  & $-$ & 4.26(31) & 2.13(6)  & 2.56(6)\\
             & 5 & $-$ & 7.03(3.21)  & 2.72(13) & 1.36(10) & 2.34(34)\\
             & 6 & 15.05(2.52) &1.47(17)  & 1.92(7)  & 1.20(1)  & 1.54(2)\\
             & 7 & 16.45(5.56) & 16.44(1.73)  & $-$ & 0.96(40)  & 2.05(1.98)\\
             & 8 & 8.70(5.95) & 3.13(16)& 0.45(6) & 0.87(2)  & 1.23(2)\\
J1709$-$4429 & 4 & 14.35(26) & 3.66(4) & $-$ & 1.93(2) & 2.30(2)\\
             & 5 & 14.03(7.12) & 3.37(6) & $-$ & 1.86(3) & 2.20(3)\\
             & 6 & 16.67(2.44) & 3.36(5) & $-$ & 1.51(2)  & 1.85(2)\\
\hline\hline
\end{tabular}
\end{table}

During the glitch, the vortices unpinned from the pinning sites of nonlinear creep region A, then move outward to the other nonlinear creep region A$^\prime$ by crossing through region B with the moment of inertia $I_\mathrm{B}$. Region B contributes to the angular momentum transfer at glitches while it cannot sustain the pinned vortices and hence cannot participate the spin-down caused by vortex creep \citep{cheng88}. According to the Eq. (\ref{conserJ}), $I_\mathrm{B}$ can be determined by 
\begin{equation}\label{IB}
    I_\mathrm{c}\Delta\Omega_\mathrm{c}=(I_\mathrm{A}/2+I_\mathrm{B}+I_\mathrm{s})\delta\Omega_\mathrm{s},
\end{equation}
where the factor 1/2 accounts for linear decrease of $\delta\Omega_\mathrm{A}$ within $I_\mathrm{A}$. The total crustal superfluid moment of inertia involved a glitch is $I_{\rm{cs}}=I_{\rm{A}}+I_{\rm{s}}+I_{\rm{B}}$.
Importantly, it has been argued that ``entrainment'' of the neutron superfluid by the crystalline structure of the crust greatly reduces its mobility, reducing the superfluid angular momentum reservoir to be tapped at glitches~\citep{2012PhRvL.109x1103A,2013PhRvL.110a1101C}. 
Due to this effect, moments of inertia of the superfluid components $I_{\rm A}$ and $I_{\rm s}$ obtained from the post-glitch timing fits should be multiplied by the same enhancement factor. 
\footnote{However, the mobility of superfluid neturons related to the effective neutron mass $m_{\rm n}^{*}/m_{\rm n}$ is an open problem which has been discussed actively in the literature, see e.g.,~\citet{2005NuPhA.747..109C,2012PhRvC..85c5801C,2017PhRvL.119f2701W,2017JLTP..189..328C}. 
More recently, it has been shown that in the so-called pasta phase, where sizeable crustal moment of inertia resides, the effect is reversed so that anti-entrainment with smaller than unity for $m_{\rm n}^{*}/m_{\rm n}$ realizes \citep{sekizawa24}. }
Regularity of glitches and data accumulated over the years have important implications on the parts of the neutron star that drive the spin-up events and thereby the moments of inertia of the corresponding regions. Thus, the long-term glitch observations can be utilized to bring constraints on the EOS. 

\section {RESULTS}
\label{sec:result}

We have analysed the timing observation data of five pulsars listed in Table \ref{Tab:publ-works} in radio and gamma-ray bands, which resulted in glitches detected in four pulsars.
In total, 12 glitches were identified, one of which was new discovery.
In Table \ref{Tab:F0F1-works}, we present the pre- and post-glitch timing solutions obtained using {\footnotesize \text{TEMPO}2}, with a parameter uncertainty of 1$\sigma$. 
The first three columns present the pulsar name, the intervals corresponding to the glitch numbers \footnote{``Glitch numbers" represent the sequence numbers of glitch events detected in a given pulsar.}, and the reference epoch (the middle moment of the timing data). The next three columns show the spin parameters obtained by fitting Eq. (\ref{equ:1}), namely $\nu$, $\dot{\nu}$ and $\ddot{\nu}$.
\footnote{Since the inter-glitch interval typically spans a long duration, it is common practice to fit $\nu$, $\dot{\nu}$ and $\ddot{\nu}$ to ensure that the timing residual remain phase-connected. Higher-order derivatives of the spin frequency are generally considered to have no practical physical significance and are regarded as merely reflecting the amplitude of timing noise.
}
The last four columns respectively summarize the number of ToAs used for fitting, the duration of observation, the data span, and the root means square (RMS) of the timing residuals.

Table \ref{Tab:glitch-works} reports the parameters of all the detected glitches.
The first three columns are the pulsar name, the reference number of each glitch, and the glitch epoch.
The middle of interval between the last pre-glitch observation time and the first post-glitch observation time is determined as the glitch epoch. 
The uncertainty is half of the interval between the two observations. 
The other parameters are obtained through fitting Eq. (\ref{equ:2}) to ToAs, while the associated uncertainty is calculated using the error propagation.
Specifically, the fourth and fifth columns show the relative change in the spin frequency ($\Delta \nu/\nu$) and the first derivative of frequency ($\Delta \dot{\nu}/\dot{\nu}$), respectively.
In the sixth column, Y represents the newly detected glitch in this work and P is the previously published glitch event.
If significant exponential recovery process is observed, we added exponential decay term to the {\footnotesize \text{TEMPO}}2 fitting, the results of these parameters ($Q$, $\tau_{\rm d}$) are shown in the seventh and eighth columns.
The last two columns are the RMS and the data span.

Besides, the evolutions of $\nu$ and $\dot{\nu}$ are shown in Figs. (\ref{1420glitch}-\ref{1709glitch}) and (\ref{1718glitch}-\ref{J1509glitch}), where values of each $\nu$ and $\dot{\nu}$ were obtained through independent fitting Eq. (\ref{equ:1})
to ToAs data spanning 150--600 d. Moreover, we applied the least squares method to perform piecewise fitting for $\dot{\nu}$ data, using the reduced chi-square value ($\chi^2_{\rm r}$) to evaluate the quality of the fit. Exponential fitting was applied to the $\dot\nu$ data within a few hundred days after the glitch, while linear fitting was applied to the $\dot\nu$ data following the exponential decay.

In the panels (d) of Figs.~(\ref{1420glitch}-\ref{1709glitch}) and (\ref{1718glitch})
, we also report the fits within the framework of the vortex creep model (see Sec.~\ref{sec:model}) (shown in green curves), done for 9 large glitches in our sample, using Eqs. (\ref{linear_response}), (\ref{nonlinear_response}) and (\ref{integ_nonlinear_response}). The corresponding fit parameters are provided in Table \ref{creepmodelfit}. 
Based on the $\chi^2_{\rm r}$ value of the fits, one sees that the recovery fit from a different and physically-motivated functional form describes the observations equally well.
Besides, various moments of inertia from the vortex-creep-model fit are collected in Table \ref{moi}. The results suggest that glitches involve only the crustal superfluid and possibly a portion of the outer core, with the adopted creep model providing sufficient capacity for the observed glitch amplitudes.

In the next section of Sec.~\ref{sec:dis}, we also show our results for the oscillations in the spin-down rates of PSRs J1718$-$3825 and J1509$-$5850, and reassess them within the vortex bending model (see purple curves in Figs.~\ref{1718glitch}-\ref{J1509glitch}). 
Moreover, beyond the common magnetic dipole-radiation scenario, we discuss the effect of the internal torque from bending vortex and inter-glitch creep contributions on the braking index ($n = \nu \ddot{\nu}/\dot{\nu}^2$). 

\subsection{PSR J1420$-$6048}

\begin{figure*}
\centering
\includegraphics[width=0.8\textwidth]{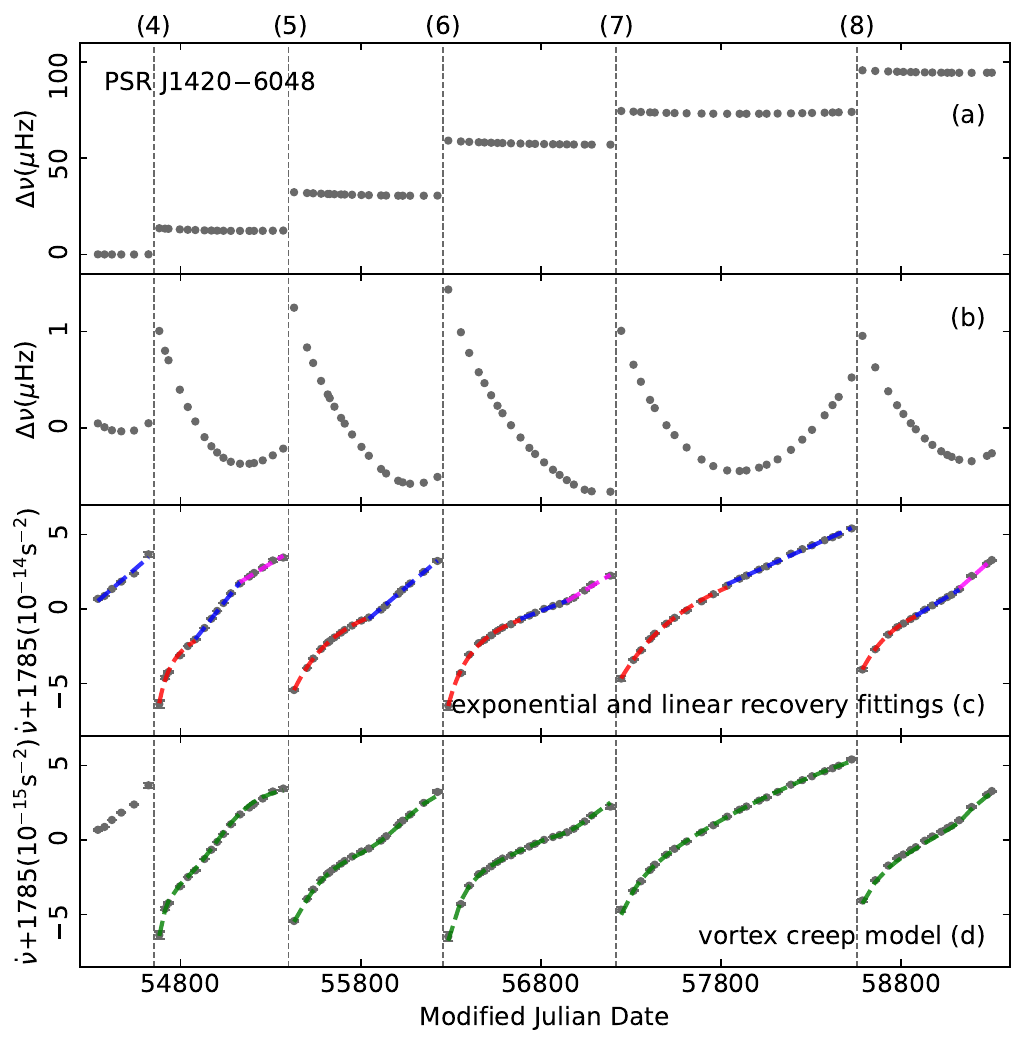}
\caption{
Glitches in PSR J1420$-$6048: The vertical gray dashed line and the numbers in between parentheses at the top represent the glitch epoch and the reference number of glitch occurrences, respectively.
Panel (a) shows the relative change ($\Delta \nu$) of spin frequency, that is, the change in pulsar rotation frequency relative to the frequency model before the glitch.
Panel (b) is an enlarged view of panel (a), showing the evolution details of $\Delta \nu$ after the glitch, which is obtained by subtracting its average value from $\Delta \nu$ of post-glitch.
Panel (c) shows the evolution of the spin-down rate ($\dot{\nu}$) over time before and after the glitch. 
The red, blue, and magenta dashed lines represent the one exponential and two different linear recovery fitting results, respectively.
In Panel (d), the green curves represent the fit using the vortex creep model (see Sec.~\ref{sec:model}), with the corresponding fit parameters provided in Table \ref{creepmodelfit}.
}
\label{1420glitch}
\end{figure*} 

PSR J1420$-$6048 is an energetic Vela-like pulsar, first discovered during the Parkes multibeam pulsar survey by \cite{D'AmicoKMC2001}.
In addition to being observed in the radio band, this young pulsar has been detected in both the X-ray and gamma-ray bands \citep{WeltevredeAAA2010}. 
X-ray pulsations (AX J1420.1$-$6049) were observed by \cite{RobertsRJ2001} using ASCA, while gamma-ray emissions (GeV J1417$-$6100/3EG J1420$-$6038) were detected by the EGRET telescope onboard the Compton Gamma Ray Observatory \citep{NishidaTAB2003}.

PSR J1420$-$6048 has a spin period of 68 ms and a period derivative of $\dot{P} \sim 83.167 \times 10^{-15}$ \citep{D'AmicoKMC2001}. From these values, the pulsar's surface magnetic field is estimated to be $B \sim 2.41 \times 10^{12}$ G, and its characteristic age is $\sim$ 13 kyr \citep{D'AmicoKMC2001}. 
This pulsar also has a high spin-down energy loss rate of $\dot{E} \sim 1.0 \times 10^{37}$ erg s$^{-1}$, placing it among the 18 most energetic known pulsars. Given its youth and large energy loss rate, PSR J1420$-$6048 is a valued
source for studying glitch activity.

To date, eight glitches have been observed in PSR J1420$-$6048 \citep{YuMHJ2013, LiuZZF2021, LowerJDS2021, LinWHT2021}, with fractional glitch sizes on the order of $\Delta \nu/\nu \sim 10^{-6}$. These glitches generally exhibit exponential recoveries, with glitch intervals of 2--3 yr. As mentioned above, \cite{LinWHT2021} found that in the gamma-ray band, the pulse profile of PSR J1420$-$6048 may change significantly before and after glitches 4 and 8.

We have combined \textit{Fermi}-LAT data with Parkes radio timing data to update the glitch epochs and parameters for glitches 4, 5, 6, 7, and 8, as shown in Table \ref{Tab:glitch-works}.
Fig. \ref{1420glitch} shows the 14 yr evolution of $\nu$ and $\dot{\nu}$ for PSR J1420$-$6048. 
Previous studies indicated no exponential decay component in the post-glitch recovery of glitch 8 \citep{LinWHT2021}. However, our analysis reveals a clear exponential recovery component, as shown in Fig. \ref{1420glitch}. We modeled the post-glitch recovery for glitch 8, finding an exponential timescale of 100(6) d and $Q = 0.0131(5)$. Additionally, we updated the fractional glitch size and its first derivative to $\Delta \nu/\nu \sim 1480(1) \times 10^{-9}$ and $\Delta \dot{\nu}/\dot{\nu} \sim 6.0(1) \times 10^{-3}$, respectively.

From panel (c) of Fig. \ref{1420glitch}, we observe that glitches 4, 6, and 8 each exhibit two distinct linear recovery phases following an initial exponential recovery. This is particularly evident for glitch 4. To verify this, we performed exponential and linear fits to the evolution of $\dot{\nu}$ (see panel (c)), and calculated the $\chi^{2}_{\rm r}$ values of these fits, which range from 0.57 to 2.25, with corresponding to degrees of freedom (d.o.f.) ranging from 2 to 11, respectively. 
Just as shown in panel (c), the linear processes of glitch 4 and glitches 6, 8 are clearly different. 
For glitch 4, the slope of the first linear recovery phase ($\ddot{\nu} \sim 1.778(3) \times10^{-21} $ s$^{-3}$) is nearly double that of the second phase ($\ddot{\nu} \sim 8.58(6) \times10^{-22}$ s$^{-3}$). In contrast, glitches 6 and 8 show the opposite trend: the slopes of the second linear phase are larger than the first. Specifically, glitch 6 has $\ddot{\nu}$ values of $5.45(2) \times 10^{-22}$ s$^{-3}$ and $8.49(5) \times 10^{-22}$ s$^{-3}$, while glitch 8 exhibits values of $8.39(2) \times 10^{-22}$ s$^{-3}$ and $1.237(6) \times 10^{-21}$ s$^{-3}$ for the two linear phases, respectively.
For glitches 5 and 7, a single linear recovery phase adequately describes the post-glitch evolution, with slopes of $1.185(2) \times 10^{-21}$ s$^{-3}$ and $6.327(9) \times 10^{-22}$ s$^{-3}$, respectively. This behaviour of two consecutive linear recovery processes in the post-glitch has not been previously reported for PSR J1420$-$6048, likely due to the limited timing data available from individual Parkes and \textit{Fermi}-LAT observations.
\footnote{Nevertheless, it should be mentioned here that the ``shoulder'' or ``elongated-S'' shape seen after the glitches in PSR J1420$-$6048 may also be due to an unaccounted part of the post-glitch recovery times which could be originated from unmodelled timing noise as discussed in~\citep{2022RPPh...85l6901A}.}
More discussions on recovery behaviour can be found below in Sec. \ref{sec:dis}.

We apply the vortex creep model to fit the five glitches of PSR J1420$-$6048, and the fit results are shown in Fig. \ref{1420glitch}, panel (d): 
\begin{itemize}
\item
G4: We fit the fourth glitch (G4) of PSR J1420$-$6048 with Eqs. (\ref{linear_response}) and (\ref{nonlinear_response}), with a $\chi^{2}_{\rm r}$ value of 1.17 (12 d.o.f.).
As seen in Table \ref{creepmodelfit}, the vortex creep model prediction for the time to the next glitch from G4 is $t_{\rm th}=698(32)$ d which is in qualitative agreement with the observed inter-glitch time $t_{\rm obs}=747(28)$ d within uncertainties.
The exponential decay time-scale obtained from the vortex creep model application $\tau_{\rm creep,exp}=41(11)$ days is slightly different than what found from fit with Eq. (\ref{equ:2}): $\tau_{\rm d}=53(7)$ d (see Table \ref{Tab:glitch-works}). 
The reason is that Eq. (\ref{nonlinear_response}) has also an exponential dependence and dominates over the initial days of the data. 
The recoupling time-scale for the crustal superfluid associated with G4 is $\tau_{\rm s}=116(10)$ d. The total fractional crustal superfluid moment of inertia involved in G4 is $I_{\rm cs}/I=2.56(6)\%$. 
\item

G5: We fit the fifth glitch (G5) of PSR J1420$-$6048 with Eqs. (\ref{nonlinear_response}) and (\ref{integ_nonlinear_response}), with the $\chi^{2}_{\rm r}$ value is 1.01 (13 d.o.f.). Due to the initial exponential dependencies of the two equations, there is no need to include a separate exponential function to fit the data of G5. 
As seen in Table \ref{creepmodelfit}, the vortex creep model prediction for the time to the next glitch from G5 is $t_{\rm th}=964(19)$ d, while the observed inter-glitch time $t_{\rm obs}=856(10)$ d. 
The origin of the discrepancy may lie in the fact that crustquake, which played the role of starting the vortex unpinning avalanche, probably left behind a tiny persistent shift in the spin-down rate by leading to a horizontal displacement of the broken platelet. Such effects on the vortex creep model estimates are also visible for some of the Crab \citep{erbil19} and the Vela \citep{akbal17} glitches. 
The recoupling time-scale for the crustal superfluid associated with G5 is $\tau_{\rm s}=109(6)$ d. The total fractional crustal superfluid moment of inertia involved in G5 is $I_{\rm cs}/I=2.34(34)\%$. Both $\tau_{\rm s}$ and $I_{\rm cs}/I$ are quite similar for G4 and G5, suggesting that these two glitches have originated from the same crustal superfluid region.
\item

G6: We fit the sixth glitch (G6) of PSR J1420$-$6048 with Eqs. (\ref{linear_response}), (\ref{nonlinear_response}) and (\ref{integ_nonlinear_response}), with the $\chi^{2}_{\rm r}$ value is 0.60 (12 d.o.f.).
As seen in Table \ref{creepmodelfit}, the exponential decay time-scale obtained from the vortex creep model application $\tau_{\rm creep,exp}=84(11)$ d is slightly different than $\tau_{\rm d}=44(8)$ d found from fit with Eq. (\ref{equ:2}) (see Table \ref{Tab:glitch-works}) again due to the initial exponential dependencies of Eqs. (\ref{nonlinear_response}) and (\ref{integ_nonlinear_response}). 
The vortex creep model prediction for the time to the next glitch from G6 is $t_{\rm th}=1124(32)$ d, while the observed inter-glitch time is $t_{\rm obs}=960(13)$ d. However, as can be clearly seen from Fig. \ref{1420glitch}, the seventh glitch (G7) arrived before the recovery of G6 has been fully completed. The recoupling time-scale for the crustal superfluid associated with G6 is $\tau_{\rm s}=98(11)$ d, similar to the value of the previous two events. The total fractional crustal superfluid moment of inertia involved in G6 is $I_{\rm cs}/I=1.54(2)\%$.
\item

G7: We fit the seventh glitch (G7) of PSR J1420$-$6048 with Eqs. (\ref{linear_response}), and (\ref{integ_nonlinear_response}), with very low $\chi^{2}_{\rm r}$ value (0.43, for 18 d.o.f.). 
As seen in Table \ref{creepmodelfit}, the exponential decay time-scale obtained from the vortex creep model application $\tau_{\rm creep,exp}=109(32)$ d is consistent with $\tau_{\rm d}=102(7)$ d found from fit with Eq. (\ref{equ:2}) (see Table \ref{Tab:glitch-works}). As can be clearly seen from Fig. \ref{1420glitch}, the post-glitch spin-down rate of G7 over-recoveries, i.e. extends over the original pre-glitch level. One possible explanation of this peculiar and rare type of relaxation in terms of the vortex creep model is as follows. According to the vortex creep model, there are two nonlinear creep regions responsible for the post-glitch rotation and spin-down rate evolution, namely vortex traps (initiating a glitch through collective vortex unpinning) with moment of inertia $I_{\rm A}$ and vortex capacitor regions (inside of which vortices cannot creep but other freed lines may pass through of it and take part in angular momentum conservation) with moment of inertia $I_{\rm B}$. If some of these capacitor regions become activated such that vortices inside that region start to creep after their interaction with the unpinned ones, then with the excess angular momentum transfer of such activated vortices the post-glitch spin-down rate will extend over the pre-glitch level.
The vortex creep model prediction for the time to the next glitch from G7 is $t_{\rm th}=938(425)$ d and agrees with the observed inter-glitch time $t_{\rm obs}=1339(14)$ d within errors. The recoupling time-scale for the crustal superfluid associated with G6 is $\tau_{\rm nl}=779(554)$ d and due to its being longer has not affected the initial exponential decay part of the glitch. The total fractional crustal superfluid moment of inertia involved in G7 is $I_{\rm cs}/I=2.05(1.98)\%$.
\item

G8: We fit the eighth glitch (G8) of PSR J1420$-$6048 with Eqs (\ref{linear_response}), (\ref{nonlinear_response}) and (\ref{integ_nonlinear_response}), and the corresponding $\chi^{2}_{\rm r}$ value of 1.87 (8 d.o.f.).
As seen in Table \ref{creepmodelfit}, the exponential decay time-scale obtained from the vortex creep model application $\tau_{\rm creep,exp}=86(39)$ d is slightly different than $\tau_{\rm d}=100(6)$ d found from fit with Eq. (\ref{equ:2}) (see Table \ref{Tab:glitch-works}) again due to the initial exponential dependencies of Eqs. (\ref{nonlinear_response}) and (\ref{integ_nonlinear_response}). 
The recoupling time-scale for the crustal superfluid associated with G8 is $\tau_{\rm s}=26(11)$ d and smallest among the glitches of PSR J1048$-$6048 analysed in this study. This value is consistent with its lower superfluid reservoir involvement in the glitch. The total fractional crustal superfluid moment of inertia involved in G8 is $I_{\rm cs}/I=1.23(2)\%$.
\end{itemize}

In addition, Eq. (\ref{nonlinear_relaxtime_erbil}) estimates that the superfluid recoupling time-scale is $\tau_{\rm nl}=$(16--763) d for PSR J1420$-$6048. This prediction matches quite well with the range $[\tau_{\rm nl}, \tau_{\rm s}]=[26(11)-779(554)]$ d obtained from the fit results given in Table \ref{creepmodelfit}.

\subsection{PSR J1028$-$5819}
\label{sec:J1028}


\begin{figure}
\centering
\includegraphics[width=0.48\textwidth]{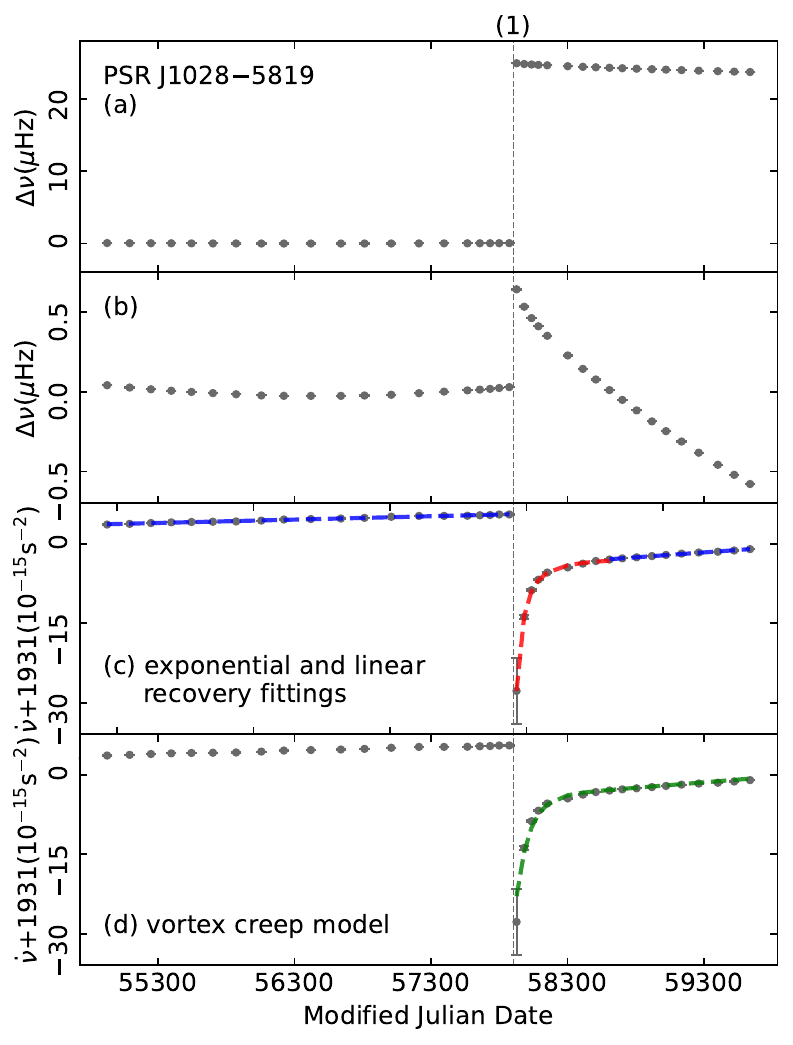}
\caption{Glitch in PSR J1028$-$5819. The descriptions of panels (a), (b), (c) and (d) are consistent with Fig. \ref{1420glitch}.
}
\label{J1028glitch}
\end{figure}

PSR J1028$-$5819 exhibits all the typical characteristics of a glitching pulsar, including a short period of $P \sim 91.4$ ms, a high period derivative of $\dot{P} \sim 16.1\times 10^{-15}$, a relatively small characteristic age of $\tau_{\rm c}$ $\sim 90$ kyr, and a large spin-down luminosity of $\dot{E} \sim 8.3\times 10^{35}$ erg s$^{-1}$ \citep{KeithJKW2008}. 
In 2009, \textit{Fermi}-LAT detected gamma-ray pulsations from PSR J1028$-$5819 for the first time \citep{AbdoAAB2009}. Later, \cite{LowerJDS2021} reported its first glitch event at MJD $\sim 57881(14)$ using Parkes timing data, with a fractional glitch size of $\Delta \nu/\nu \sim 2296.5(5) \times 10^{-9}$.

We have combined 14 yr of \textit{Fermi}-LAT data with 11 yr of Parkes timing data.
As shown in Fig. \ref{J1028glitch}, a glitch occurred at MJD $\sim 57904(6)$. This event was a large glitch (G1) with a fractional size of $\Delta \nu/\nu \sim 2284.0(9) \times 10^{-9}$, corresponding to a frequency increment of $\Delta \nu \sim 24.99(1)$ $\upmu$Hz. 
Our data are denser than those of \cite{LowerJDS2021}, allowing for a more precise fitting of the exponential decay, which results in a slightly weaker glitch amplitude.
 The evolution of $\dot{\nu}$, shown in Fig. \ref{J1028glitch} panel (c), reveals an exponential recovery process following the glitch. The fitted time constant for this exponential decay is $\tau_{\rm d} = 62(7)$ d, with $Q = 0.0060(2)$. This $Q$ value is half that of the glitch at MJD $\sim 57881(14)$ \citep{LowerJDS2021}.
For this pulsar (and later for PSR J1718$-$3825 as well), we also analysed the timing noise using the ``Cholesky'' method, given that the time spans exceeded three years\footnote{Since timing noise is characterized by low-frequency noise, commonly referred to as red noise and has a long time scale of years, it is typically analyzed over a time span of three years or more.}.
For PSR J1028$-$5819, we found a spectral index ($\alpha$) of 4 both before and after the glitch (see below in Sec. \ref{sec:n} for further dicsuccsions).

Large glitches with exponential recovery are often accompanied by long-term linear recovery, which continues until the next glitch event occurs \citep{YuMHJ2013}. The large glitch in PSR J1028$-$5819 follows this trend. As shown in panel (c) of Fig. \ref{J1028glitch}, the blue and red dashed lines represent the fitting curve of $\dot{\nu}$ linear and exponential decay, respectively.
Among them, the $\chi^2_{\rm r}$ values of linear fitting in pre- and post-glitch remain consistent with 1.06, corresponding to d.o.f. of 18 and 8, respectively.
Furthermore, the  $\chi^2_{\rm r}$ value for the exponential fitting of post-gltch is 1.60, corresponding to a d.o.f. of 5. 
We found that the post-glitch slope ($\ddot{\nu} \sim 2.14(5)\times 10^{-23}$ s$^{-3}$) is nearly three times larger than the pre-glitch slope ($\ddot{\nu} \sim 7.35(14) \times 10^{-24}$ s$^{-3}$), indicating that the glitch caused a permanent change in the spin-down rate.

We fit this glitch with Eqs. (\ref{linear_response}) and (\ref{integ_nonlinear_response}), and the fit result is shown in Fig. \ref{J1028glitch}.
The $\chi^2_{\rm r}$ value of this fitting is 0.22, corresponding to a d.o.f. of 13.
The total fractional crustal superfluid moment of inertia involved in the glitch is estimated to be $I_{\rm cs}/I = 3.35(16)\%$, as seen in Table \ref{moi}.
The vortex creep model predicts that the time to the next glitch following G1 is $t_{\rm th} = 4890(269)$ d. Since not all of the angular momentum reservoir within the crustal superfluid was depleted during G1, PSR J1028$-$5819 may experience another glitch due to a different pinned superfluid layer before the recovery from the previous event is fully complete. The fit result for the crustal superfluid's recoupling timescale, $\tau_{\rm nl} = 71$ d, is close to the lower end of the theoretical estimate range $\tau_{\rm nl} =$ 108--5113 d, as derived from Eq. (\ref{nonlinear_relaxtime_erbil}).

%
\begin{figure*}
\centering
\vspace{-0.3cm}
\includegraphics[width=0.7\textwidth]{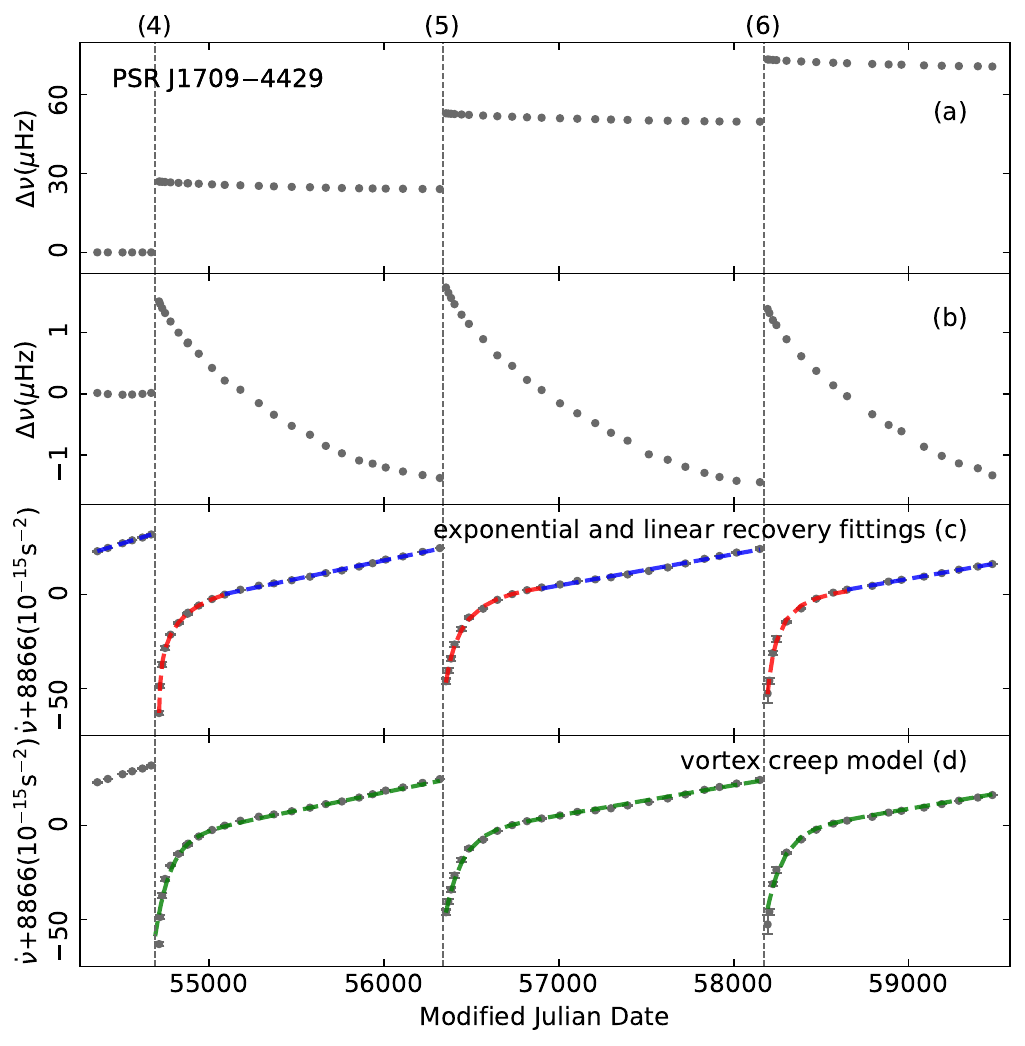}
\caption{{Glitches in PSR J1709$-$4429. The descriptions of panels (a), (b), (c) and (d) are consistent with Fig. \ref{1420glitch}.
In panel (d), the red dashed lines are the fitting result of the exponential recovery process of post-glitch 4, 5 and 6, and the number of fitting exponential are 2, 1 and 1, respectively.
}
}
\label{1709glitch}
\end{figure*}
\begin{figure}
\centering
\vspace{-0.5cm}
\includegraphics[width=1.0\linewidth]{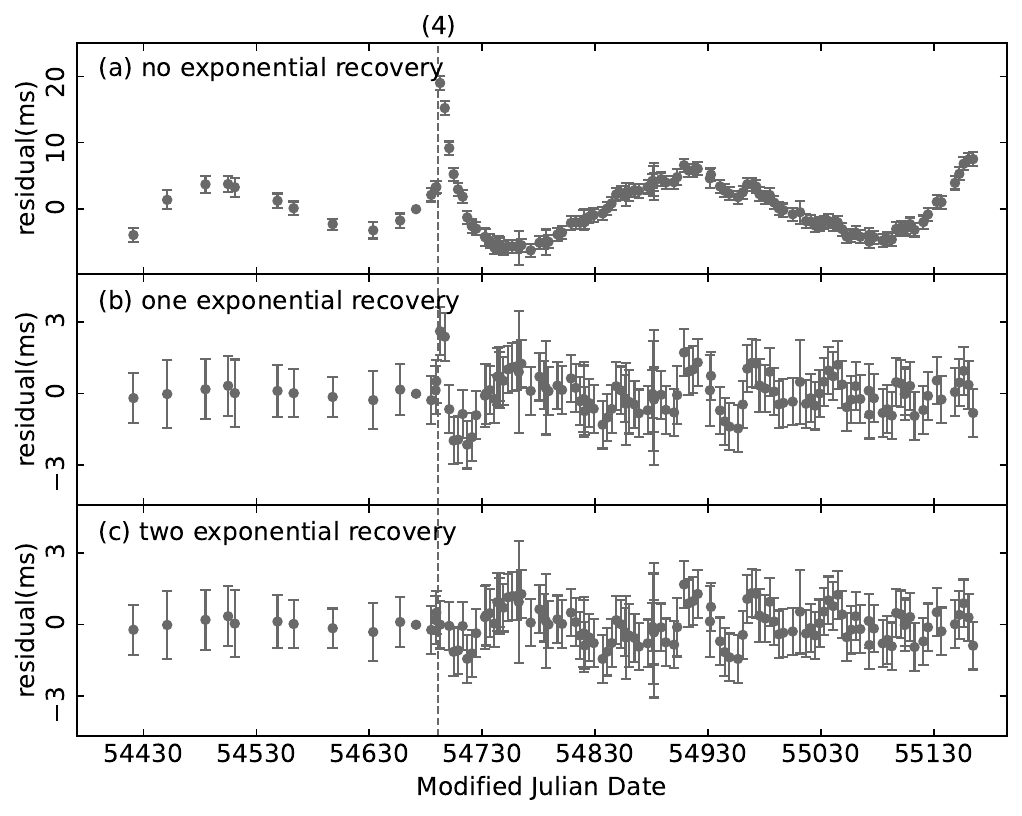}
\caption{ PSR J1709$-$4429 timing residuals of glitch 4. The three panels represent the timing residuals for the spin-down model with 0, 1, and 2 exponential decay terms, respectively. The vertical gray dashed line and the numbers in between parentheses at the top represent the glitch epoch and the reference number of glitch occurrences, respectively.
 }
\label{1709fit}
\end{figure}

\subsection{PSR J1709$-$4429 (B1706$-$44)}





 In 1992, the pulsed radio emission from PSR J1709$-$4429 was discovered in the Parkes high-frequency survey project \citep{JohnstonSMK1992}.
\cite{ThompsonBBE1996} detected gamma-ray pulsations with EGRET, and Chandra found X-ray pulsed radiation in 2002 \citep{GotthelfHD2002}.
 This is a bright Vela-like pulsar with a period of 102.5 ms, characteristic age of 17.5 kyr and the spin-down energy loss rate of $\dot{E} \sim 3.4 \times 10^{36}$ erg s$^{-1}$ \citep{WangMPB2000}.
 In 1995, the first glitch of PSR J1709$-$4429 was reported \citep{JohnstonMLK1995}, and five more glitches were successively published in the following timing observations \citep{WeltevredeAAA2010, YuMHJ2013, LowerFBB2018, LowerJDS2021}.

In Fig. \ref{1709glitch}, we present the evolution of $\nu$ and $\dot{\nu}$ over the last 14 yr. Panels (a) and (b) clearly show that three glitches occurred during this period (glitches 4, 5, and 6). 
These events are all large glitches, with fractional glitch sizes of $\Delta \nu/\nu \sim 10^{-6}$, corresponding to frequency increments of approximately 25 $\upmu$Hz. 
For glitch 4, \cite{YuMHJ2013} reported that post-glitch behaviour has only one exponential recovery.
However, when we perform a fit combining \textit{Fermi}-LAT and Parkes timing data, two exponential decay characteristics are revealed in the timing residuals, as shown in Fig. \ref{1709fit}.
It is evident that higher-cadence observations following the glitch benefit uncovering the decay behavior in detail.
The first exponential decay feature is {$Q$1 = 0.0104(5) and $\tau_{\rm d1}$ = 72(4)} d, and the second exponential decay feature is {$Q$2 = 0.006(1) and $\tau_{\rm d2}$ = 4.2(6)} d.
The transient increment in the frequency and frequency first derivative is {$\Delta \nu/\nu \sim 2777(4) \times 10^{-9}$} and {$\Delta \dot{\nu}/\dot{\nu} \sim 64(15) \times 10^{-3}$}, respectively.
For glitches 5 and 6, the glitch parameters obtained by our fitting are consistent with the previously reported results within the error range, and the detailed results are shown in Table \ref{Tab:glitch-works}.
Moreover, it is noticeable that there are differences in the cadence of ToAs before and after the glitch. The observational cadence prior to the glitch was lower, primarily because the \textit{Fermi}-LAT did not collect data during this period (\textit{Fermi}-LAT began operations in 2008). To assess the impact of varying observation cadences on the timing results, we used the \texttt{FAKE} plug-in of {\footnotesize \text{TEMPO}}2 to simulate multiple sets of ToA data with different cadences for glitch 4. The simulation showed that the sparser observations produced results consistent with those from the denser observations for this glitch event.

Moreover, we performed exponential and linear fits to the evolution of $\dot{\nu}$, with the results displayed as red and blue dashed lines, respectively, in panel (c). 
These fitted $\chi_{\rm r}^{2}$ values range from 0.32 to 1.16, corresponding to a d.o.f. range of 4 to 12.
For glitch 4, both single and double exponential functions were used to model the post-glitch exponential recovery of $\dot{\nu}$. 
By fitting the exponential recovery of $\dot{\nu}$, we found that glitch 4 achieved the highest accuracy of fit ($\chi_{\rm r}^2 = 0.32$, for 4 d.o.f.) with a double exponential model, compared to $\chi_{\rm r}^2 = 1.37$ (7 d.o.f.) for a single exponential fit. The result is consistent with that of the two exponential recovery procedures observed in the data processing. 
For glitches 5 and 6, an exponential term accurately describes the recovery of $\dot{\nu}$, with $\chi_{\rm r}^2$ values close to 1, specifically 0.81 (7 d.o.f.) and 1.12 (5 d.o.f.).
Interestingly, although the linear recovery slopes change significantly after glitches 1, 2, and 3 \citep{YuMHJ2013}, our linear fitting results show that the linear recovery process after glitches 4, 5, and 6 exhibits a similar slope, which is approximately $\ddot{\nu} \sim  2.0 \times10^{-22}$ s$^{-3}$.

We apply the vortex creep model to fit the three glitches, using Eqs. (\ref{linear_response}) and (\ref{integ_nonlinear_response}), and the $\chi_{\rm r}^2$ values are 1.09 (19 d.o.f.), 0.86 (18 d.o.f.), and 0.64 (12 d.o.f.) in order. 
The fit result is shown in Fig. \ref{1709glitch} panel (d): 
\begin{itemize}
\item
G4: The exponential decay time-scale obtained from the vortex creep model application $\tau_{\rm creep,exp}=86(9)$ d is slightly different than $\tau_{\rm d}=72(4)$ d found from fit with Eq. (\ref{equ:2}) (see Table \ref{Tab:glitch-works}) again due to the initial exponential dependency of Eq. (\ref{integ_nonlinear_response}). 
The vortex creep model prediction for the time to the next glitch from G4 is $t_{\rm th}=1680(16)$ d and in qualitative agreement with the observed inter-glitch time $t_{\rm obs}=1648(4)$ d. The total fractional crustal superfluid moment of inertia involved in G4 after corrected for the crustal entrainment effect is $I_{\rm cs}/I=2.30(2)\%$. The recoupling time-scale for the crustal superfluid associated with G4 is $\tau_{\rm nl}=3.4$ d.
\item
G5: The exponential decay time-scale obtained from the vortex creep model application $\tau_{\rm creep,exp}=95(19)$ d is slightly different than $\tau_{\rm d}=61(6)$ d found from fit with Eq. (\ref{equ:2}) again due to the initial exponential dependency of Eq. (\ref{integ_nonlinear_response}). The vortex creep model prediction for the time to the next glitch from G5 is $t_{\rm th}=1861(25)$ d and matches with the observed inter-glitch time $t_{\rm obs}=1836(4)$ d quite well. The total fractional crustal superfluid moment of inertia involved in G5 is $I_{\rm cs}/I=2.20(3)\%$. The recoupling time-scale for the crustal superfluid associated with G4 is $\tau_{\rm nl}=28$ d.
\item
G6: The exponential decay time-scale obtained from the vortex creep model application $\tau_{\rm creep,exp}=93(9)$ d differs from $\tau_{\rm d}=47(8)$ d obtained through fit with Eq. (\ref{equ:2}) again due to the initial exponential dependency of Eq. (\ref{integ_nonlinear_response}). The vortex creep model prediction for the time to the next glitch from G6 is $t_{\rm th}=1861(25)$ d, similar to the its previous glitch. The total fractional crustal superfluid moment of inertia involved in G6 is $I_{\rm cs}/I=1.85(2)\%$. The recoupling time-scale for the crustal superfluid associated with G6 is $\tau_{\rm nl}=3.2$ d.
\end{itemize}

\subsection{PSR J1718$-$3825}

\begin{figure}
\centering
\includegraphics[width=0.49\textwidth]{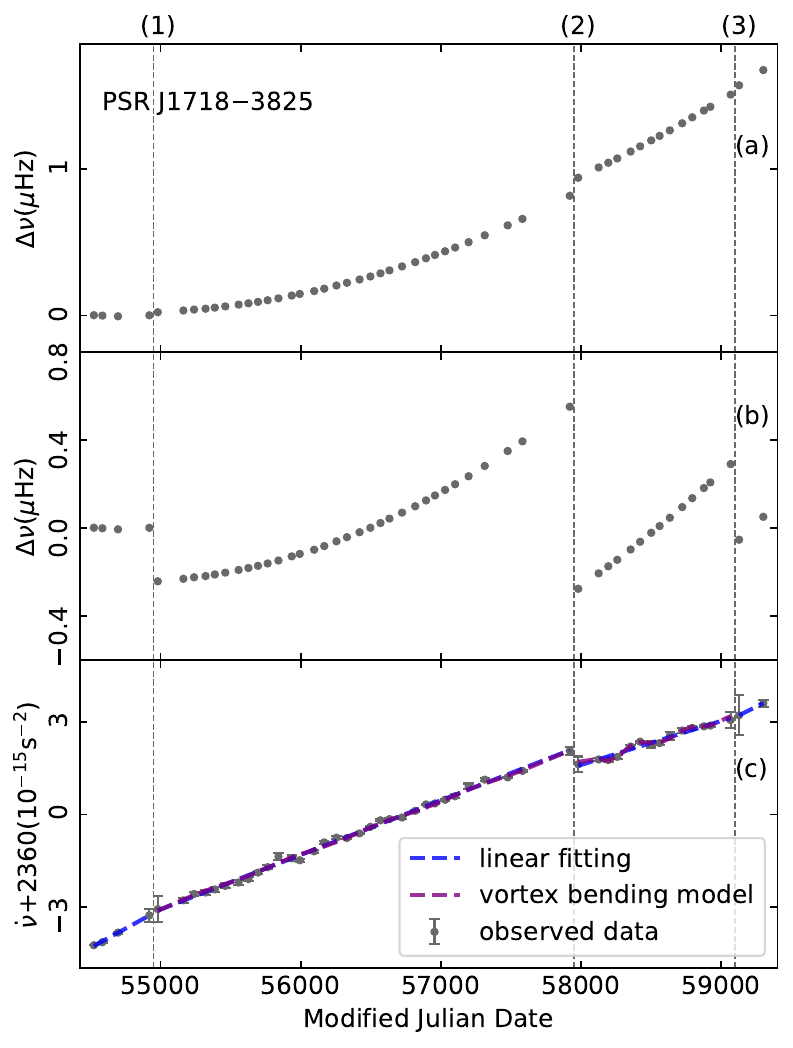}
\caption{
Glitches in PSR J1718$-$3825. 
The descriptions of panels (a), (b), and (c) are consistent with Fig. \ref{1420glitch}. 
In Panel (c), the blue dashed curves represent the linear fitting of the linear recovery process at pre- and post-glitch.
The purple curves show the fit with the vortex bending model (see Sec.~\ref{sec:dis}).
}
\label{1718glitch}
\end{figure}

PSR J1718$-$3825 is a young pulsar, discovered by \cite{ManchesterLCB2001} in the radio band. It has a characteristic age of 89.5 kyr, a spin period of 74.7 ms, and a relatively high spin-down luminosity of $\dot{E} \sim 1.3 \times 10^{36}$ erg s$^{-1}$. In addition to an associated X-ray nebula \citep{HintonCGK2007}, it is also a source of gamma-ray emission (HESS J1718) within the supernova remnant RX J1713.7$-$3946 \citep{AharonianABB2007}.

We analysed more than 14 yr of timing data for PSR J1718$-$3825 in the radio and gamma-ray bands.
A total of 3 glitches were identified, of which glitch 1 and 2 have been reported in the radio band \citep{YuMHJ2013,LowerJDS2021}, and glitch 3 is a new small glitch.
By examining the timing residuals, it was deduced that glitch 3 occurred at MJD $\sim$ 59121(8).
Using the fitting method, we determine the {fractional} glitch size to be {$\Delta \nu/\nu \sim 1.9(2) \times 10^{-9}$} and the first frequency derivative to be {$\Delta \dot{\nu}/\dot{\nu} \sim -0.12(9) \times  10^{-3}$}.
We also analyzed the timing noise using the pre- and post-glitch timing data of glitch 2. It was found that the spectral index remained consistently at 6 before and after the glitch.

In Fig. \ref{1718glitch}, the three detected glitches in PSR J1718$-$3825 are presented. It is clear that $\dot{\nu}$ only experienced a long-term linear recovery after the glitches, and there may be stable $\ddot{\nu}$.
Therefore, based on the glitch epochs, the $\dot{\nu}$ data were divided into four intervals. 
and linear fits to $\dot\nu$ data were applied to each interval to determine $\ddot{\nu}$. 
The $\chi_{\rm r}^{2}$ values of these linear fits were 1.54, 0.66, and 0.74, respectively, with corresponding d.o.f. of 2, 30, and 12. 
The fits revealed that, under the influence of glitch, $\ddot{\nu}$ has undergone obvious changes, with values of $2.9(1) \times 10^{-23}$ s$^{-3}$, $2.04(2)\times 10^{-23}$ s$^{-3}$, and $1.65(9) \times 10^{-23}$ s$^{-3}$, respectively.
These fitted values are represented by the blue dashed lines in panel (c) of Fig. \ref{1718glitch}.
A comparison with predictions from the superfluid vortex bending model is also provided for the first (G1) and second (G2) glitch events. Detailed discussions are included below in Sec.~\ref{sec:dis}. 


\subsection{PSR J1509$-$5850}

\begin{figure}
\centering
\includegraphics[width=1.0\linewidth]{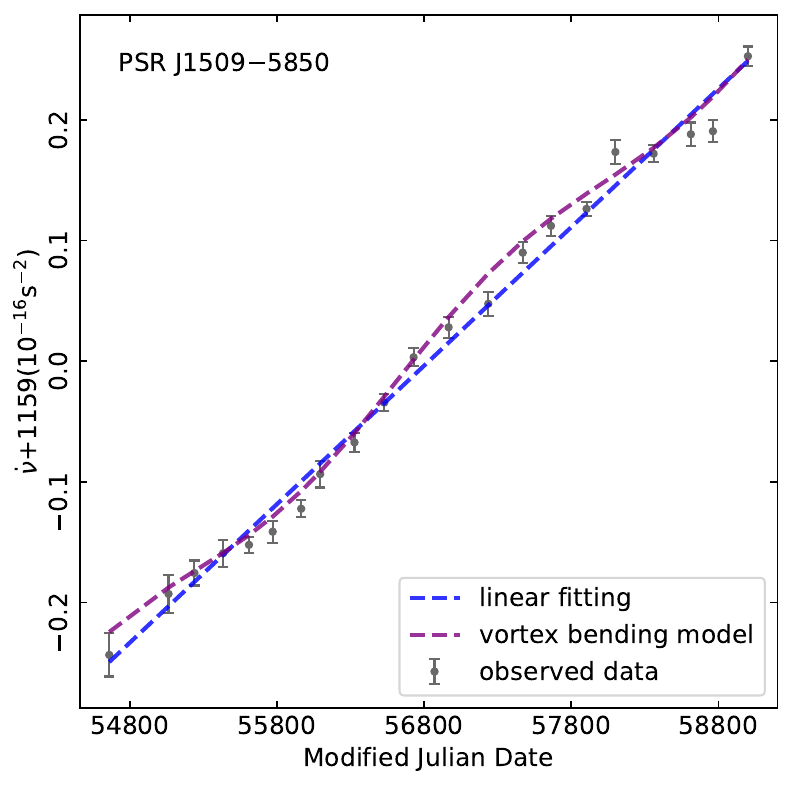}
\caption{Evolution of spin-down rate over time for PSR J1509$-$5850, along with a comparison between a linear fit (blue) and the vortex bending model (purple) describing well the oscillation, as discussed in Sec.~\ref{sec:dis}.
}
\label{J1509glitch}
\end{figure}

PSR J1509$-$5850 is a middle-aged pulsar with a characteristic age of 154 kyr \citep{KramerBML2003}. We have analysed nearly 14 yr of timing data in both the radio and gamma-ray bands and found no glitches within the observed data range. Our study of the spin-down rate evolution suggests that PSR J1509$-$5850 exhibits a roughly stable $\ddot{\nu}$, as shown in Fig. \ref{J1509glitch}. Using the least squares method for linear fitting of $\dot\nu$ data (yielding an $\chi_{\rm r}^{2}$ value of 0.80 with 19 d.o.f.), we determined $\ddot{\nu}$ to be $1.33(3) \times 10^{-24}$ s$^{-3}$, with the fitted result indicated by the blue dotted line in Fig. \ref{J1509glitch}. Additionally, the figure presents a fit based on the superfluid vortex bending model, which describes well the $\dot{\nu}$'s oscillation (see Sec.~\ref{sec:dis}). 

\section {DISCUSSION }
\label{sec:dis}

\subsection{Oscillations in the spin-down rate}
\label{sec:bending}

\begin{table*}
\caption{Fit results for the application of the vortex bending model. See text for details.
}\label{bendingmodelfit}
    \setlength{\tabcolsep}{30pt}      
        \renewcommand{\arraystretch}{1.15}
\begin{tabular}{lccc}
\hline  \hline
Parameter                                                         & J1509$-$5850                              & J1718$-$3825  G1                     & J1718$-$3825  G2         \\
\hline 
Data Range (MJD)   &54303--59510   & 54997--57940   & 57964--59113            \\
$A$ (10$^{-10}$\,\rm s$^{-2}$)                    &  6.86(41)                       & 2.43(58)                       & 10.95(26)            \\
$\tau$ (d)                                                    & 5928                                         & 619090                                    & 218                                 \\
$\Omega_{0}$ (10$^{-8}$ rad\,s$^{-1}$)     & 2.63(27)                       & 11.07(1.53)                      & 19.94(56)           \\
$\phi_{\rm{b}}$ (rad)                                                       &0.20(53)                    & 2.53(2.32)                            &1.50(2)           \\
$\dot\nu_{0}$ (10$^{-12}$\,\rm s$^{-2}$)   &    $-1.15926(1)$     & $-2.36321(5)$          & $-2.35852(2)$ \\
$\ddot\nu_{0}$ (10$^{-24}$\,\rm s$^{-3}$) & 1.34(4)           &20.25(32)            &16.71(45)    \\
$T_{\rm osc}$ (yr)                                            &7.567                                       &1.799                                           &0.998                                \\                                     
$\ell$ (10$^{5}$\,cm)                          &3.72                                               &0.98                                                  &0.54                                        \\
\hline \hline
\end{tabular}
\end{table*}

Commonly, the observed spin evolution of pulsars is simply attributed to the emission of magnetic dipole radiation and ejection of charge particles from the magnetosphere (pulsar wind; \citealt{manchester1977}).
Nevertheless, inside the pulsars, vortex lines could experience bending due to pinning in the inner crust and outer core, and applies torque on the surface of a neutron star as a result of their time variable coupling to the magnetospheres after transient events like mode switching, giant pulses, pulsar wind ejection, radiative bursts and glitches. 
Combining the linear creep response, the vortex bending results in an internal torque with vortex-bending-assisted oscillation modes and a damped sinusoidal-like oscillations in the star's spin-down rate evolution \citep{erbil2023}. 

After considering the negative braking index due to magnetic field change, the long-term spin-down properties are expressed as follows \citep{erbil2023}:
\begin{equation}
    \nu(t)=A\exp\left(-\frac{t}{2\tau}\right)\sin\left(\Omega_0t+\phi_{\rm{b}}\right)+\left(\dot{\nu}_0+\frac{1}{2}\ddot{\nu}_0t\right)t \ ,
\end{equation} \label{eq:bendingmodel_Omega}
and
\begin{equation}
    \begin{aligned}\dot{\nu}(t)=&A\Omega_{0}\exp\left(-\frac{t}{2\tau}\right)\cos\left(\Omega_{0}t+\phi_{\rm{b}}\right)\\&-\frac{A}{2\tau}\exp\left(-\frac{t}{2\tau}\right)\sin\left(\Omega_{0}t+\phi_{\rm{b}}\right)\\&+\dot{\nu}_{0}+ \ddot{\nu}_{0}t \ .
    \end{aligned} \label{eq:bendingmodel_Omegadot}
\end{equation}
Here, $A$, $\Omega_0$ and $\phi_{\rm{b}}$ are the amplitude, frequency and phase of the oscillations arising from the internal torque contribution due to vortex bending. $\dot\nu_{0}$ and $\ddot\nu_{0}$ are the spin-down rate and the second time derivative of the spin frequency of the neutron star due to external torques acting on its surface. The damping time of the oscillations in the spin-down rate $\tau$ is related to the linear creep recoupling time $\tau_{\rm l}$ given by Eq. (\ref{linear_relaxtime}) as $\tau=(I_{\rm c}/I)\tau_{\rm l}$. Here, $I$ is the total moment of inertia of the neutron star. $I_{\rm c}$ is the sum of the moment of inertia of the neutron star normal matter and core superfluid/superconducting components that couples to its surface on short timescales and tied by magnetic field crossing the whole star. Basically, $I_{\rm c}$ excludes the inner crust superfluid and the toroidal field region in the outer core of neutron stars. Oscillation time is defined as
\begin{equation}
    T_{\rm osc}=\frac{2\pi}{\Omega_{0}}=\frac{2\pi}{\omega_{0}}\left[1-\left(\frac{1}{2\tau\omega_{0}}\right)^{2}\right]^{-1/2}\ ,
\end{equation}
with
\begin{equation}
    \omega_{0}=\left(\frac{I_{\rm b}I}{I_{\rm cs}^{\prime}I_{\rm c}}\frac{10\Omega_{\rm c}\kappa}{\pi\ell^{2}}\right)^{1/2}\ .
\end{equation}
Here, $I_{\rm cs}^{\prime}$ is the moment of inertia of the crustal superfluid, $I_{\rm b}$ is the moment of inertia of the regions in which vortex lines has bent over due to pinning (crustal superfluid plus toroidal field region in the outer core) with $\ell$ being the length-scale over which vortices bent and $\kappa$ is the vorticity quantum attached to each line.

We apply the vortex bending model to the timing solution of PSR J1509$-$5850 between MJD 54303--59510, obtaining a fit result with a $\chi_{\rm r}^{2}$ value of 0.87 (15 d.o.f.), see Fig. \ref{J1509glitch}. This result represents a significant improvement compared to the linear fit provided earlier in Sec.~4.5.
The model free parameters corresponding to the vortex bending model's application, along with the oscillation time $T_{\mathrm{osc}}$ and length scale $\ell$ are given in Table \ref{bendingmodelfit}. 
If we attribute the large frequency second time derivative of $\ddot\nu\cong1.37\times10^{-24}$ \,s$^{-3}$ to an unresolved Vela-like glitch occurred before the time interval we considered, then Eq. (\ref{tig}) with the typical value $\Delta\dot\nu_{\rm p}/\dot\nu=5\times10^{-3}$ gives the estimate of $t_{\rm ig}=158$ yr for the time between two successive giant glitches hypothesised to be experienced by this pulsar. 
Thus, we predict that PSR J1509$-$5850 will glitch again within the same timescale. Future observations of this pulsar are required to test our assertion.

We also analyse the timing data of PSR J1718$-$3825 between MJD 54303--59491. Since there is not enough data for the third glitch (G3), we made an application of the vortex bending model to its first (G1) and second (G2) glitch events. 
See Fig. \ref{1718glitch}, panel (c). 
The fit parameters corresponding to the vortex bending model are given in Table \ref{bendingmodelfit}.
The $\chi_{\rm r}^{2}$ values for the fitting results of G1 and G2 are 0.76 (26 d.o.f.) and 0.17 (8 d.o.f.), respectively.
We see that, while the vortex bending model slightly improves the fit to the data of G1 compared to the linear one (provided above in Sec.~4.4), it provides an excellent fit to the G2 data reducing the residuals significantly over the linear one.
In particular, the damping timescale of the oscillations in the spin-down rate is well constrained to be $\tau=218$ d. When assessed with Eq. (\ref{linear_relaxtime}), this would provide important aspects for the microphysical parameters related to nuclear pinning mechanism as well as temperature estimate for the neutron star crust. Also, the length-scale $\ell$ found from the fits would provide a new method to put constraints on the crustal thickness and in turn on the EOS of the neutron stars. It should be also important to address these constraints in light of the sinusoidal-like oscillations observed in the spin-down rates of various pulsars.

\subsection{Inter-glitch Braking Index}
\label{sec:n}

\begin{table} 
\caption{The braking indices $n_{\rm{ig}}$ predicted in the vortex model. Also shown is the comparison of the observed inter-glitch $\ddot{\nu}_{\rm obs}$ with the vortex creep model prediction $\ddot{\nu}_{\rm ig}$ given by Eq. (\ref{nuddotint}). }\label{nuddotcomp}
   \renewcommand{\arraystretch}{1.15}
    \setlength{\tabcolsep}{4pt}      
\begin{tabular}{cccccc}
\hline     \hline 
PSR &Gl. No. & Epoch & $n_{\rm{ig}}$ & $\ddot{\nu}_{\rm ig}$   & $\ddot{\nu}_{\rm obs}$  \\
 &{ }     & (MJD) &  &(10$^{-24}$s$^{-3}$) & (10$^{-24}$s$^{-3}$) \\ 
\hline  
J1028$-$5819  &1  & 57904(6) & 67(4) &22.9(13) &20(3) \\
J1420$-$6048  &4  & 54653(19) & 96.8(26) &2097(56) &1145(26) \\
              &5  & 55400(9) & 59(6) &1281(139) &932(36) \\
              &6  & 56256(1) &44.2(23) &958(50) &1172(59) \\
              &7  & 57216(12) & 30.6(9) &664(20) &779(9) \\
              &8  & 58555(2) & 54.1(11) &1170(24) &551(11) \\
J1709$-$4429  &4  & 54691(2) &42.8(5) &343(4)  &351(4) \\
              &5  & 56339(2) & 23(4) &188(31)  &135(91) \\
              &6  & 58175(2) & 35(4) &282(30)  &227(43) \\
\hline  \hline 
\end{tabular}
\end{table}

If only the external torque is considered, based on the use of torque laws of the form $-K\Omega^n$, where $K$ is a constant and the braking index is assumed to be 3 as predicted by the magnetic dipole model~\citep{manchester1977}.
Following the vortex creep model \citep{AlparAS1984,erbil17}, glitches may occur as a result of collective unpinning of a very large number of vortex lines in the crustal superfluid and inter-glitch spin evolution of a pulsar is dominated by the response of the star to the variations in the superfluid internal torques' coupling due to such vortex discharges. 
In terms of the model parameters theoretical expectation of the inter-glitch braking index can be recast as \citep{AlparBA2006}:
\begin{equation}
    n_{\rm{ig}}=(\beta +1/2)\left[\left(\Delta\dot\nu_{\rm p}/\dot\nu\right)_{-3}^{2}/(\Delta\nu_{\rm p}/\nu)_{-6}\right] \ ,
\end{equation}
where the glitch magnitudes in the rotation and the spin-down rates are expressed in units of $10^{-6}$ and $10^{-3}$, respectively. Also, we have used the shorthand notation $\beta=I_{\rm B}/I_{\rm A}$. As a fiducial value, in our calculations below we take $\beta=4.11$, which was obtained for PSR J1048--5832 \citep{LiuYGY2024}.
In comparison, the $constant$ braking index ($n$) before and after the glitch were estimated according to the fitted $\ddot{\nu}$ and the parameters ${\nu}$, $\dot{\nu}$ in Table \ref{Tab:F0F1-works}.
We observed stable values of $\ddot{\nu}$ for PSRs J1028$-$5819, J1509$-$5850 and J1718$-$3825 and fitted the slopes of $\ddot{\nu}$ for each.
In the case of PSR J1028$-$5819, the braking indices before and after the glitch are 21.7(4) and 63(1), respectively. 
For glitch 2 of PSR J1718$-$3825, the braking indices before and after the glitch are 49.2(5) and 40(2), respectively. 
Although we did not identify a glitch event for PSR J1509$-$5850, its $\dot{\nu}$ has steadily evolved over 14 yr,  resulting in a braking index of $n = 11.1(3)$. 
These non-canonical braking indices may be indicative of internal torque at work, likely associated with the dynamics of nucleon superfluidity in this context.
Furthermore, PSR J1028$-$5819 exhibits a significant change in the braking index before and after the glitch, while maintaining a consistent spectral index ($\alpha$) of 4 in both periods (see Sec. \ref{sec:J1028}). This suggests that the change in the braking index is unlikely to be related to red noise.

In addition, the most prominent effect of vortex unpinning on the spin parameters is $\ddot\nu$ and given by
\begin{equation}
    \ddot\nu_{\rm ig}=\left(\beta+\frac{1}{2}\right)\left(\frac{\Delta\dot\nu_{\rm p}}{\dot\nu}\right)^{2}_{-3}\left(\frac{\Delta\nu_{\rm p}}{\nu}\right)^{-1}_{-6}\left(\frac{\dot\nu^{2}}{\nu}\right) \ .
    \label{nuddotint}
\end{equation}
The inter-glitch time is expressed as
\begin{equation}
    t_{\rm ig}=\frac{\Delta\dot\nu_{\rm p}}{\ddot\nu_{\rm ig}}=2\times10^{-3}\left(\beta+\frac{1}{2}\right)\tau_{\rm c}\left(\frac{\Delta\dot\nu_{\rm p}}{\dot\nu}\right)^{-1}_{-3}\left(\frac{\Delta\nu_{\rm p}}{\nu}\right)_{-6} \ .
    \label{tig}
\end{equation}
Comparison of the prediction of the vortex creep model [Eq. (\ref{nuddotint})] with observed inter-glitch $\ddot{\nu}$ for our sample of glitches is given in Table \ref{nuddotcomp}. An inspection to the Table \ref{nuddotcomp} reveals that the agreement between the vortex creep model predictions and observations are qualitatively good except for G8 of PSR J1420$-$6048. The reason for the discrepancy may be a persistent step increase that not resolved which leads to a somewhat larger estimate in $\Delta\dot\nu_{\rm p}$. Future observations will help to test this hypothesis.

\section{CONCLUSIONS}

In this study, we combined the \textit{Fermi}-LAT and the Parkes timing data of gamma-ray pulsars J1028$-$5819, J1420$-$6048, J1509$-$5850, J1709$-$4429 and J1718$-$3825 for more than 14 yr to study glitches. 
Our analysis has led to the identification of a total of 12 glitch events in four pulsars, among which there is one new glitch event. These glitches exhibit fractional sizes $\Delta \nu/\nu$ ranging from $1.9(2)\times 10^{-9}$ to $2964(4) \times 10^{-9}$. 
Furthermore, we have undertaken a comprehensive update of glitch-related parameters for glitches previously reported. 
An intriguing aspect of our study is the effective combination of the ToAs from both \textit{Fermi}-LAT and Parkes, which not only augments the effective observation cadence of these pulsars, 
but more importantly, have the potential to fill the observational gap after the glitch occurs, thereby accurately revealing the exponential recovery process on the short time-scale of post-glitch.
As a result, our investigation has unveiled several novel phenomena within the glitch recovery process.

In the case of glitch events  4, 6, and 8 of PSR J1420$-$6048, we discovered two linear recovery processes in the post-glitch evolution of $\dot{\nu}$.
Furthermore, regarding glitch 8, we noted the presence of an exponential recovery process with $Q = 0.0131(5)$, $\tau_{\rm d} = 100 (6)$ d.
For PSR J1709$-$4429, we detected two exponential decay processes within glitch 4, with $Q1=0.0104(5)$, $\tau_{\rm d1}=72(4)$ d, $Q2=0.006(1)$, $\tau_{\rm d2}=4.2(6)$ d, and the parameter $\Delta \dot{\nu}/\dot{\nu}$ exhibited significant difference, 
standing at $64(15) \times 10^{-3}$.
Meanwhile, we found that $\ddot{\nu}$ changed significantly during post-glitch linear recovery in glitches 1, 2 and 3, but not in glitches 4, 5 and 6.
For PSR J1718$-$3825, a new small glitch was identified in MJD $\sim$ 59121(8), with glitch parameters are $\Delta \nu/\nu \sim 1.9(2) \times 10^{-9}$ and $\Delta \dot{\nu}/\dot{\nu} \sim -0.12(9) \times  10^{-3}$.

We connected the observed glitches with the dynamics of neutron vortex arrays inside the pulsars and discussed the theoretical interpretation of these events in the framework of vortex creep model. 
For our sample of glitches, application of the vortex creep model revealed that the total fractional moment of inertia of the crustal superfluid required to tapped at glitches is $\lesssim3.5\%$, 
indicating that only crustal superfluid and probably some part of the outer core region of a neutron star are involved in glitches. 
Yet another implication of the application of the vortex creep model is that the model estimate for the time to the next glitch in general agrees well with the observed inter-glitch time for a given pulsar's particular glitch.  
Thus, by doing timing fit to a particular glitch future observations can be planned for corresponding pulsar to establish any variations in the pulse profile concurrent with the glitch occurrence. Such a detection will enable us to connect the magnetospheric physics to the interior dynamics and disclose the interplay between them. 
Different parts of the post-glitch recovery, in particular the exponential relaxation can be studied to infer microphysics related to the neutron star crustal and core matter. 
We plan to compute the relaxation time derived from microscopic two- and three-body nuclear forces, incorporating pasta phases and proton effective masses self-consistently,  as well as application to the observed cases in a separate paper~\citep{tu2024}.
We'd also like to address the effects of more updated superfluidity/superconductivity-related interaction parameters by taking proximity etc. into account on the exponential recovery time estimate.
Simulating pulsar glitches through experimental observations in rotating ultracold atoms offers exciting potential to constrain the complex physics associated with superfluid vortices~\citep{2023PhRvL.131v3401P}.
Future astrophysical observations will provide a larger sample, particularly with variations in post-glitch behaviour, promising greater precision in glitch constraints and offering insights into the dynamics of superfluid vortices beyond the coarse-grained two-component scenario.
Additionally, pulsar glitches may imprint themselves on gravitational wave radiation, establishing a promising link between multi-messenger studies of these phenomena and insights into the dense matter EOS~\citep{2024MNRAS.528.1360C,2024PhRvD.109h3006W,2024MNRAS.532.3893Y}.

\section*{Acknowledgements}
This work is supported by National SKA Program of China (no.~2020SKA0120300), the Strategic Priority Research Program of the Chinese Academy of Sciences (grant no. XDB0550300), the National Natural Science Foundation of China (grant nos 12494572, 12273028, 12041304 and 12288102), and the Major Science
and Technology Program of Xinjiang Uygur Autonomous Region (grant no. 2022A03013, 2022A03013-4).
EG acknowledges support from the National Natural Science Foundation of China (NSFC) programme 11988101 under the foreign talents grant QN2023061004L.
ZZ is supported by the Natural Science Basic Research Program of Shaanxi (program no. 2024JC-YBQN-0036).
SD is supported by the Guizhou Provincial Science and Technology Foundation (no. ZK[2022]304) and the Scientific Research Project of the Guizhou Provincial Education (no. KY[2022]132).
PW acknowledges support from the National Natural Science Foundation of China under grant U2031117 and the Youth Innovation Promotion Association CAS (id. 2021055).
DL is supported by the 2020 project of Xinjiang Uygur autonomous region of China for flexibly fetching in upscale talents.
We acknowledge the use of the public data from the \textit{Fermi}-LAT data archive.
The Parkes radio telescope is part of the Australia Telescope National Facility which is funded by the Commonwealth of Australia for operation as a National Facility managed by CSIRO. This paper includes archived data obtained through the CSIRO Data Access Portal.

\section*{Data Availability}

The data underlying this article will be shared on reasonable request to the corresponding authors.

\bibliographystyle{mnras}
\bibliography{glitch4}

\end{document}